\documentclass[pre,twocolumn,preprintnumbers,amsmath,amssymb,nopacs,nofootinbib,floatfix]{revtex4}

\usepackage{graphicx,bm,xcolor, bbold}
\usepackage[normalem]{ulem}
\usepackage{hyperref}

\makeatletter
\def\graphicscale{\twocolumn@sw{0.3}{0.4}}
\def\graphicthreescale{\twocolumn@sw{0.3}{0.4}}

\begin{document}

\title{Critical crossover phenomena driven by symmetry-breaking defects\\at quantum transitions}

\author{Alessio Franchi}
\affiliation{Dipartimento di Fisica dell'Universit\`a di Pisa
        and INFN, Largo Pontecorvo 3, I-56127 Pisa, Italy}

\author{Davide Rossini}
\affiliation{Dipartimento di Fisica dell'Universit\`a di Pisa
        and INFN, Largo Pontecorvo 3, I-56127 Pisa, Italy}

\author{Ettore Vicari}
\altaffiliation{Authors are listed in alphabetic order.}
\affiliation{Dipartimento di Fisica dell'Universit\`a di Pisa
        and INFN, Largo Pontecorvo 3, I-56127 Pisa, Italy}

\date{\today}

\begin{abstract}
We study the effects of symmetry-breaking defects at continuous
quantum transitions (CQTs), which may arise from localized external
fields coupled to the order-parameter operator.  The problem is
addressed within renormalization-group (RG) and finite-size scaling
frameworks. We consider the paradigmatic
one-dimensional quantum Ising models at their CQT, in the presence of
defects which break the global ${\mathbb Z}_2$ symmetry.  We show that
such defects can give rise to notable critical crossover
regimes where the ground-state properties experience substantial and
rapid changes, from symmetric conditions to 
symmetry-breaking boundaries.  An effective characterization of these
crossover phenomena driven by defects is achieved by analyzing the
ground-state fidelity associated with small changes of the defect
strength.  Within the critical crossover regime, the fidelity
susceptibility shows a power-law divergence when increasing the system
size, related to the RG dimension of the defect strength; in contrast,
outside the critical defect regime, it remains finite.  We support
the RG scaling arguments with numerical results.
\end{abstract}

\maketitle


\section{Introduction}
\label{intro}

Critical phenomena have attracted a lot of interest
in the last decades (see,
e.g., Refs.~\cite{Wilson-nobel,Fisher-74,ZJ-book,Cardy-book,SGCS-97,
  Sachdev-book, DMS-book,PV-02,NO-11,RV-21} and references
therein). One of the reasons is that their emerging features have a
great degree of universality, being largely independent of the
microscopic details. Therefore they have a wide applicability to
different systems and within very different physical contexts.
Moreover, they allow us to describe complex phenomena using a
relatively small number of relevant variables, providing a notable
simplification of the analysis of many-body systems.  However,
critical phenomena occur under particular conditions, when the
system develops long-range correlations, for example at continuous
phase transitions arising from thermal or quantum fluctuations.

Some important features of critical phenomena at thermal and quantum
continuous phase transitions are related to the presence of
boundaries~\cite{Binder-DL,Barber-DL,Diehl-DL,AL-91,CL-91, CZ-94,SM-12}
and to perturbations arising from the presence
of defects~\cite{Brown-82,OA-96,OA-97,CKPPS-00,FMV-05,EP-10,CMV-12,CKM-21}.
These are not
academic issues, since physical systems have generally boundaries and
are subject to localized defects of various nature.  The presence of
isolated defects does not generally change the bulk power-law
behaviors characterizing the critical behavior of observables at large
scale. However, their effects may get somehow amplified by the
long-range critical modes at continuous transitions, in the
neighborhood of the defect, and in particular in finite-size systems.

In this paper we investigate the effects of symmetry-breaking defects
in quantum many-body systems at continuous quantum transitions (CQTs).
We address these issues exploiting renormalization-group (RG) and
finite-size scaling (FSS) frameworks.  We argue that, although the
presence of isolated defects does not generally change the bulk
power-law behaviors at CQTs, they can drive notable critical crossover
behaviors when the defects are the only source of symmetry
breaking. In this case they induce critical crossovers in the
ground-state and low-energy properties, between limiting cases that
can be associated with different boundary conditions: from boundary
conditions (or absence of boundaries) preserving the global symmetry
to boundary conditions breaking the symmetry associated with the
CQT.  Therefore, the addition of isolated symmetry-breaking defects to
critical (strictly symmetric) systems can give rise to substantial
changes of the ground states, and the finite-size behavior of the
critical modes, even in the large-size limit within the FSS regime
around the CQT.  Two different situations, which develop different
RG properties, must be distinguished: whether the defects are located
within the bulk of the system, or at the boundaries. 

We challenge this general scenario within the paradigmatic
one-dimensional quantum Ising systems in a transverse field, studying
the effects of local defects breaking the global ${\mathbb Z}_2$
symmetry, within the bulk and at the boundaries.  We analyze the
crossover behaviors induced by the defects at its CQT, when varying
their strength.  These critical defect crossovers provide a bridge
between situations that can be associated with different boundary
conditions: from translation-invariant FSS behaviors in Ising rings to
FSS behaviors of systems with parallel fixed boundary conditions
(PFBC). The power laws characterizing these critical crossovers are
determined by the RG dimensions of the perturbations arising from the
defects, which differ for defects within the bulk and at the
boundaries.  The scaling theory that we develop is then checked by
numerical computations.

An effective characterization of the crossover phenomena driven by
symmetry-breaking defects is obtained by analyzing the ground-state
fidelity measuring the overlap between ground states associated with
different defect parameters.  This provides information on the
variations of the ground-state structures due to the defect,
whether it gives rise to substantial changes involving the
whole system, or the changes remain limited to a finite region. We
argue that the susceptibility associated with the defect fidelity
diverges in the large-size limit within the critical crossover regime,
while it remains finite outside it.  Such a power-law divergence is
related to the RG dimension of the defect parameter.

The paper is organized as follows.  In Sec.~\ref{models} we introduce
the models that we consider, i.e., quantum Ising rings without
boundaries [corresponding to periodic boundary conditions (PBC)] in
the presence of one symmetry-breaking defect, and Ising chains with
boundaries [such as open boundary conditions (OBC)] with defects
localized at the boundaries.  In Sec.~\ref{obser} we introduce the
quantities that we use to monitor the effects of the symmetry-breaking
defects, including the ground-state fidelity associated with
small changes of the defect parameters. In Sec.~\ref{critcross} we
outline the description of the critical defect crossover phenomena in
Ising rings, using RG and FSS frameworks. In Sec.~\ref{fssboufield} we
discuss the case of Ising chains with symmetric boundaries, when we
add symmetry-breaking boundary defects.  In Sec.~\ref{numres} we
present numerical analyses supporting the scaling behaviors obtained
by the RG and FSS analyses.  Finally, in Sec.~\ref{conclu} we
summarize and draw our conclusions.

\section{Quantum Ising models with symmetry-breaking defects}
\label{models}

The quantum Ising chain is a useful theoretical laboratory where
fundamental issues of quantum many-body systems can be throughly
investigated, exploiting the exact knowledge of several features of its
phase diagram and quantum correlations. Many results for its
low-energy properties have been derived in the ordered and
disordered phases, and in particular at the quantum critical point
separating the two phases, in the thermodynamic limit and in the FSS
limit with various boundary conditions (see, e.g.,
Refs.~\cite{Sachdev-book,DMS-book,NO-11,RV-21} and references therein).

In our study of critical crossover behaviors driven by symmetry-breaking
defects, we consider quantum Ising chains with ring-like geometry
without boundaries and in the presence of one defect, and chains with
boundaries, such as OBC, in the presence of defects localized at the boundaries.

\subsection{Quantum Ising rings with defects}
\label{rings}

Quantum Ising rings are defined by the Hamiltonian
\begin{equation}
  \hat H_r = - J \sum_{x=1}^{L} \hat \sigma^{(1)}_{x\phantom{1}}
  \hat \sigma^{(1)}_{x+1} - g \sum_{x=1}^L \hat \sigma^{(3)}_x \,,
\label{Iring}
\end{equation}
where $L$ is the system size, $\hat \sigma^{(i)}_x$ are the Pauli matrices
on the $x$th site ($i = 1,2,3$ labels the three spatial directions) and
$\hat \sigma^{(i)}_{L+1} = \hat \sigma^{(i)}_1$, corresponding to PBC.
In the following we assume
ferromagnetic nearest-neighbor interactions with $J=1$.

The model undergoes a CQT at $g=g_c=1$, belonging to the
two-dimensional Ising universality class, separating a disordered
phase ($g>g_c$) from an ordered ($g<g_c$) one (see, e.g.,
Refs.~\cite{Sachdev-book,RV-21}).  Approaching the CQT, the system
develops long-distance correlations, with length scales $\xi$
diverging as $\xi \sim |g-g_c|^{-\nu}$ where $\nu=y_g^{-1}=1$ and
$y_g$ is the RG dimension associated with the difference $g-g_c$. The
ground-state energy gap gets suppressed as $\Delta \sim \xi^{-z}$
where $z$ is the dynamic
critical exponent $z=1$. Another independent critical exponent arises
from the RG dimension of the symmetry-breaking homogeneous longitudinal
field $h$ coupled to $\sum_x \hat \sigma_x^{(1)}$, which is
$y_h=(2+d+z-\eta)/2=2-\eta/2$ where $d$ stands for the system
dimension (here $d=1$) and $\eta=1/4$, thus $y_h=15/8$. We
recall that, along the $|g|<1$ line, the longitudinal field $h$ drives
quantum first-order transitions.  Around $g_c$, the interplay between
$\xi$ and the size $L$ of the system gives rise to
FSS~\cite{Privman-90,RV-21}, defined as the large-$L$ limit keeping
$\xi/L$ constant.

We want to study the effects of localized defects breaking the
${\mathbb Z}_2$ symmetry of model~\eqref{Iring}, such as the one
described by the Hamiltonian term
\begin{equation}
  \hat D_k = - \kappa \, \hat \sigma_k^{(1)}
\label{dxdef}
\end{equation}
and localized on site $k$. Such a defect
also breaks the translation invariance of the original model~\eqref{Iring}.
Its effects within the first-order transition line for
$g<g_c$ has been analyzed in Refs.~\cite{CPV-15-tr,PRV-18}, where it
gives rise to a defect-driven CQT between different quantum phases. In
the following, we focus on the critical crossover phenomena driven by
$\hat D_k$ at the CQT for $g\approx g_c$.

We note that the defect~\eqref{dxdef} provides a bridge between
translation-invariant systems with PBC for $\kappa=0$,
cf. Eq.~\eqref{Iring}, and models with PFBC when $\kappa\to\infty$
(due to the fact that the state at site $k$ gets fixed to
the eigenstate of $\hat \sigma_k^{(1)}$ with eigenvalue $s=1$).
Note that,
in the presence of $n>1$ equal defects like that in Eq.~(\ref{dxdef}),
the limit $\kappa\to\infty$ gives rise to an effective multipartition
of the system, where the $n$ subsystems separated by the defects can
be considered as effectively disconnected chains with PFBC.

The effects of the local perturbation arising from defect may get
amplified by long-distance correlations at CQTs.  Although they do not
alter the leading power-law behavior, scaling functions may acquire a
nontrivial dependence on the external localized field, i.e., the
parameter $\kappa$ in Eq.~\eqref{dxdef}.  Indeed, as we shall see, one
symmetry-breaking defect gives rise to a critical crossover behavior
entailing substantial and rapid changes of the ground-state properties.

\subsection{Quantum Ising chains with boundary defects}
\label{chains}

We also consider another class of symmetry-breaking
defects, localized at the boundaries of the model. As we shall
see, they give rise to similar critical crossover effects at quantum
transitions, but characterized by different critical exponents.
Quantum Ising chains with OBC are defined by the Hamiltonian
\begin{equation}
  \hat H_b = - J\sum_{x=1}^{L-1} \hat \sigma^{(1)}_{x\phantom{1}} \hat \sigma^{(1)}_{x+1}
  - g \sum_{x=1}^L \hat \sigma^{(3)}_x \,.
  \label{Isc}
\end{equation}
As before, we fix $J=1$.
We discuss the effects of longitudinal
fields localized at the boundaries, such as those described by the
Hamiltonian term
\begin{equation}
  \hat B = - \zeta \left(\hat \sigma_1^{(1)} + \hat \sigma_L^{(1)}\right)\,,
  \label{bdefbou}
\end{equation}
where $\zeta$ plays the role of parallel boundary field.  This kind 
of defects allows us to interpolate between systems with OBC at
$\zeta=0$ and systems with PFBC in the limit $\zeta\to\infty$.  The
effects of boundary fields such as that in Eq.~\eqref{bdefbou} have
been already discussed in Ref.~\cite{CPV-15}.  We will add 
further results, to characterize the critical crossover that they
give rise.

\section{Observables}
\label{obser}

\subsection{Gap, magnetization and two-point function}
\label{magtwo}

We define the gap $\Delta$ as the energy difference
between the first excited state and the ground state. We
recall that the power law of the asymptotic finite-size behavior of
the gap is not changed by the presence of defects, or by different
choices of the boundary conditions.  However, the amplitude of the
leading behavior does depend on these features. For example, at the
critical point one has
\begin{equation}
  \Delta(L) = \frac{C_\Delta}{L} + O(L^{-2})\,, \label{deltafss}
\end{equation}
with $C_\Delta=\pi/2,\,\pi,\,4\pi$, respectively for PBC, OBC, and
PFBC~\cite{CPV-14, CPV-15, BG-85, CJ-86, BC-87}.

We also address the expectation value of the longitudinal
order-parameter operator $\hat \sigma_x^{(1)}$ on the ground state
$|\Psi_0\rangle$, i.e., the magnetization,
\begin{equation}
  M_x = \langle \Psi_0|\, \hat \sigma_{x}^{(1)} \, |\Psi_0\rangle\,,
  \label{magdef}
\end{equation}
its two-point correlation function
\begin{equation}
  G(x,y) \equiv \langle \Psi_0|\,\hat \sigma_{x}^{(1)} \hat \sigma_{y}^{(1)}\,
  |\Psi_0\rangle\,,\label{gxy}
\end{equation}
as well as the corresponding susceptibility and correlation length,
\begin{equation}
  \chi_x = \sum_y G(x,y)\,,\qquad
    \xi_x^2 = { \sum_y D(x,y)^2 G(x,y) \over 2 \chi_x}\,,
\label{chixidef}
\end{equation}
where $D(x,y)$ is the minimum distance between the sites $x$ and $y$
(this definition takes into account the ring geometry).  In our
analyses of position-dependent observables, such as $M_x$, $\chi_x$
and $\xi_x$, we consider particular values of $x$, such as the central
site for chains with boundaries. For Ising rings
with defect, one may choose either the site $x=k$ of the defect or the
opposite site $x=k+L/2$ at the largest distance.

We also study the spatially averaged quantities
\begin{equation}
  M_a = {1\over L} \sum_x M_x\,, \qquad
  \chi_a = {1\over L} \sum_x \chi_x \,,
  \label{avedef1}
\end{equation}
and
\begin{equation}
  \qquad \xi_a^2 =
  { \sum_{x,y} D(x,y)^2 G(x,y) \over 2 \sum_x \chi_x}\,.
\label{avedef}
\end{equation}

\subsection{RG invariant quantities}
\label{rginv}

To characterize the crossover behavior due to the defects, we consider
a set of RG invariant quantities, which we will generically denote with
$R$ in the following.  They are the ratios between the correlation
lengths and the size, i.e.,
\begin{equation}
R_{\xi x}\equiv \xi_x/L\,,\qquad R_{\xi a} \equiv \xi_a/L \,.
\label{rxidef}
\end{equation}
Moreover we may consider ratios of correlation function at different
scaling distance, such as
\begin{equation}
  R_{G x} \equiv {G(x,x+X_1 L)\over G(x,x+X_2L)}\,,\label{rgdef}
\end{equation}  
where $X_1$ and $X_2$ are fixed fractions of the size $L$, such as
$X_1=1/2$ and $X_2=1/4$.  A natural choice for Ising chains is that of
identifying $x$ with the center of the chain. For Ising rings one may
also consider the average definition
\begin{equation}
R_{G a}\equiv\frac{\sum_xG(x, x+X_1L)}{\sum_x G(x, x+X_2L)}\,.
\label{rgadef}
\end{equation}

We also consider the so-called Binder parameter depending on one point $x$,
\begin{equation}
  U_x = { \sum_{u,v,w} \langle \Psi_0 | \hat \sigma_x^{(1)} \hat \sigma_u^{(1)}
    \hat \sigma_v^{(1)} \hat \sigma_w^{(1)} |\Psi_0 \rangle \over L \left( \sum_u
    \langle \Psi_0 | \hat \sigma_x^{(1)} \hat \sigma_u^{(1)} |\Psi_0 \rangle \right)^2} \,,
\label{uxdef}
\end{equation}
or its spatially averaged version
\begin{equation}
  U_a = { \sum_{x,u,v,w}\langle \Psi_0 | \hat \sigma_x^{(1)} \hat \sigma_u^{(1)}
    \hat \sigma_v^{(1)} \hat \sigma_w^{(1)} |\Psi_0 \rangle \over \left( \sum_{x,u}
    \langle \Psi_0 | \hat \sigma_x^{(1)} \hat \sigma_u^{(1)} |\Psi_0 \rangle \right)^2}\,.
\label{uxdefave}
\end{equation}

\subsection{Ground-state fidelity associated with the defect}
\label{qinfo}

To characterize the effects of the defects, we may also consider the
ground-state fidelity quantifying the overlap between the
ground states for different defect parameters (see, e.g.,
Refs.~\cite{AFOV-08, Gu-10, BAB-17,RV-21}).  The usefulness of the
fidelity as a tool to distinguish quantum states can be
traced back to Anderson's orthogonality
catastrophe~\cite{Anderson-67}: the overlap of two many-body ground
states corresponding to Hamiltonians differing by a small perturbation
vanishes in the thermodynamic limit.  Besides that, the corresponding
fidelity susceptibility covers a central role in quantum estimation
theory~\cite{BC-94, Paris-09}, being proportional to the so-called
quantum Fisher information.  The latter indeed quantifies the inverse
of the smallest variance in the estimation of the varying parameter,
such that, in proximity of quantum transitions, metrological
performances are believed to drastically improve~\cite{ZPC-08, IKCP-08}.

To monitor the changes of the ground-state wave function
$|\Psi_0(L,g,\kappa)\rangle$ when varying the defect strength $\kappa$
by a small amount $\delta \kappa$ (keeping the Hamiltonian parameter
$g$ fixed), we define the fidelity
\begin{equation}
  A(L,g,\kappa,\kappa+\delta\kappa) \equiv \big| \langle
  \Psi_0(L,g,\kappa+\delta\kappa) | \Psi_0(L,g,\kappa)\rangle \big| \,.
  \label{fiddef}
\end{equation}
Assuming $\delta\kappa$ sufficiently small, one can expand
Eq.~\eqref{fiddef} in powers of $\delta\kappa$:
\begin{equation}
  A(L,g,\kappa,\delta\kappa) = 1 - \tfrac12 \delta\kappa^2 
  A_2(L,g,\kappa) + O(\delta\kappa^3) \,,
  \label{expfide}
\end{equation}
where
\begin{equation}
  A_2(L,g,\kappa) = {\partial^2 A
    \over \partial \, \delta\kappa^2} \bigg|_{\delta\kappa=0}\,,
  \label{a2def}
\end{equation}
represents the fidelity susceptibility~\cite{Gu-10,RV-21}.
The cancellation of the linear term in the expansion~\eqref{expfide}
is essentially related to the fact that the fidelity
is bounded~\cite{Gu-10}, i.e., $1\ge A \ge 0$.
One may also consider analogous definitions for the model~\eqref{Isc}
with the boundary fields~\eqref{bdefbou}, by replacing $\kappa$
with $\zeta$ in the above equations.

\section{Critical crossover driven by symmetry-breaking defects}
\label{critcross}

We start to consider the case of the Ising ring with one
symmetry-breaking defect [cf., Eq.~\eqref{dxdef}], whose location is
irrelevant, since the original model~\eqref{Iring} is invariant under
translations.

\subsection{Critical crossover behavior driven by the defect}
\label{rgweight}

We first show that the perturbation arising from the defect is
relevant in critical Ising rings, i.e., the RG dimension $y_\kappa$ of
the corresponding parameter $\kappa$ is positive. One can
straightforwardly determine $y_\kappa$ by analyzing the corresponding
perturbation to the translationally invariant field theory.
This can be written as
\begin{equation}
  P_\kappa = \kappa \int d\tau \,\varphi(x,\tau)\,,
\label{pft}
\end{equation}
where $\varphi(x,\tau)$ is the order-parameter field of the Ising
transition, and the integration is over the euclidean time $\tau$.
There is no integration over the space, because the perturbation is
spatially localized (note that the location of the defect is assumed
to be within the bulk, i.e., it is not close to a boundary).  The
standard RG analysis of the perturbation $P_\kappa$ entails
a relation among the RG dimensions of the quantities entering
its definition, given by
\begin{equation}
  y_\kappa + y_\varphi-z=0\,,
  \label{rgrel}
  \end{equation}
where
\begin{equation}
  y_\varphi = (d+z-2+\eta)/2 = 1/8\,
  \label{yvarphi}
\end{equation}
is the RG dimension of the order-parameter
field~\cite{Sachdev-book,RV-21}.  Thus we obtain
\begin{equation}
  y_\kappa = z - y_\varphi = 7/8\,.
\label{ykappa}
\end{equation}

We are now in the position to put forward the scaling behaviors of the
various observables introduced in Sec.~\ref{obser}, within a standard FSS
framework~\cite{Binder-DL,Barber-DL,Diehl-DL,CL-91,CZ-94,SM-12,CPV-14,RV-21}.
Since the RG dimension of the defect parameter $\kappa$ is positive,
the defect gives rise to a relevant perturbation, whose effect is that
of moving away from the FSS behavior of the Ising ring (\ref{Iring}),
characterized by translation invariance and intact ${\mathbb Z}_2$
symmetry.  The corresponding FSS limit of generic observables is
defined as the large-size limit keeping the scaling variables
\begin{equation}
  W = (g-g_c) L^{y_g}\,,\qquad
  K  = \kappa L^{y_\kappa}\,, \label{UKdef}
\end{equation}
fixed (we recall that $y_g=\nu^{-1}=1$ for one-dimensional quantum
Ising models).  This defines the critical crossover regime driven by
the defect, which develops around $\kappa=0$, for $|\kappa|\sim
L^{-y_\kappa}$.

\subsection{Critical crossover behavior of the observables}
\label{fssobs}

The critical crossover behaviors of the various quantities introduced
in Sec.~\ref{obser} can be put forward as follows.
The gap is expected to behave as
\begin{equation}
  \Delta(L,g,\kappa) \approx L^{-z} {\cal D}(W,K)\,,
  \label{gapfss}
\end{equation}
where ${\cal D}(W,K)$ is a universal scaling function, i.e..
microscopic variations of the Ising ring Hamiltonian, for example
adding further next-to-nearest neighbor couplings between the spin
operators, do not change it (apart from trivial normalizations of the
arguments).

The magnetization in the FSS limit is expected to behave as
\begin{subequations}
\begin{eqnarray}
  M_x(x,L,g,\kappa) &\approx& L^{-y_\varphi} {\cal
    M}_x(X_k,W,K)\,,
  \qquad  \label{mxfss}\\
  M_a(L,g,\kappa) &\approx& L^{-y_\varphi} {\cal M}_a(W,K)\,,
  \label{mfss}
\end{eqnarray}
\end{subequations}
where $y_\varphi$ is the RG dimension of the order-parameter field,
cf. Eq.~(\ref{yvarphi}), and
\begin{equation}
  X_k = x_k/L\,, \quad x_k = {\rm Min}[x-k,L-x+k]\,, \label{Xk}
  \end{equation}
where $x_k$ is the distance from the defect along the ring.
Analogously, for two-point correlation functions we find
\begin{equation}
  G(x,y,L,g,\kappa) \approx L^{-2y_\varphi} {\cal
    G}(X_k,Y_k,W,K)\,,\quad
  \label{gxyfss}
\end{equation}
where $Y_k=y_k/L$, and $y_k$ is defined analogously to $x_k$.  Using
the above results, we may easily derive the FSS behaviors of the RG
invariant quantities $R$ defined in Eqs.~(\ref{rxidef}), (\ref{rgdef})
and (\ref{rgadef}), obtaining
\begin{subequations}
\begin{eqnarray}
  R_x(x,L,g,\kappa) &\approx& {\cal R}_x(X_k,W,K)\,,
  \label{rxfss}\\
  R_a(L,g,\kappa) &\approx& {\cal R}_a(W,K)\,.
  \label{Rfss}
\end{eqnarray}
\end{subequations}
Analogous scaling behaviors are expected for the Binder parameters
$U_x$ and $U_a$, cf.~Eqs.~\eqref{uxdef} and~\eqref{uxdefave}.

We remark that the above FSS behaviors are defined in the large-$L$
limit keeping all arguments of the scaling functions fixed, and in
particular $K=\kappa L^{y_\kappa}$.  They reflect the fact that the
perturbation arising from the symmetry-breaking defect is relevant,
thus affecting, and changing, the asymptotic translation-invariant FSS
functions of the original Ising ring. We stress again that, unlike
quantum first-order transitions~\cite{CNPV-14,PRV-18-fofss}, they
cannot change the bulk power laws at CQTs, but only the FSS
functions~\cite{RV-21}.

As already mentioned in the previous sections, the $\kappa\to\infty$
limit of the defect strength gives rise to systems with PFBC.
Therefore we may interpret the above-reported FSS behaviors as a
critical crossover from the FSS of the translation-invariant Ising
ring to that of systems with PFBC, which is asymptotically realized
for any finite value of $\kappa>0$.  Therefore, we expect that the
limit $K\to\infty$ of the scaling functions converges to the FSS
functions for PFBC, which are realized for any finite $\kappa>0$
asymptotically in the large-$L$ limit.

In particular, the critical gap should behave as
\begin{equation}
  \Delta(L,g,\kappa>0) \approx L^{-z} {\cal D}_{\rm pfbc}(W)\,,
  \label{dlpfbc}
\end{equation}
for any finite $\kappa>0$, where ${\cal D}_{\rm pfbc}(W)$ is the scaling
function associated with the gap of Ising chains with PFBC, 
\begin{equation}
  \Delta_{\rm pfbc}(L,g) \approx L^{-z} {\cal D}_{\rm pfbc}(W)\,,\quad
    {\cal D}_{\rm pfbc}(0) = 4\pi\,,
\label{pfbcsca}
\end{equation}
and we also used Eq.~(\ref{deltafss}).
Moreover, we expect that
\begin{equation}
  {\cal D}(W,K\to\infty) = {\cal D}_{\rm pfbc}(W)\,.
  \label{infkbeh}
\end{equation}
Analogous considerations
apply to the other observables. For example, in the case of the RG
invariant quantities $R_a$, whose scaling is reported in
Eq.~\eqref{Rfss},
\begin{equation}
  R_a(L,g,\kappa>0) \approx {\cal R}_{\rm pfbc}(W) = {\cal R}(W,K\to\infty)\,,
\label{Rfsskagt0}
\end{equation}
when taking the FSS limit, keeping $\kappa>0$ fixed.

We finally remark that the critical crossover behaviors driven by
symmetry-breaking defects may appear analogous to those arising from
relevant perturbations at a unstable fixed point of the homogeneous
theory, driving away the RG flow toward another stable fixed point.

\subsection{Scaling corrections}
\label{scalco}

The asymptotic FSS behaviors are generally approached with power-law
suppressed corrections~\cite{CPV-14,RV-21}, which may depend on the
observable considered.
Within CQTs belonging to the two-dimensional Ising universality class,
the contributions of the leading irrelevant
operator are suppressed as $L^{-\omega}$ with
$\omega=2$~\cite{CPV-14,CHPV-02,CCCPV-00}.  Moreover the leading
corrections related to the breaking of the rotational invariance on
the lattice are suppressed as $L^{-\omega_{\rm nr}}$ with
$\omega_{\rm nr}=2$, as well~\cite{CHPV-02,CPRV-98,CH-00}.
However there are also corrections that are suppressed more slowly.
Some quantities are subject to corrections from analytic background
contributions~\cite{CPV-14}, for example those involving second-moment
correlation lengths, cf. Eq.~\eqref{rxidef}, for which the leading
scaling corrections get suppressed as $L^{-3/4}$ only, even in the
case of systems without boundaries.

The existence of the defect, breaking translation invariance, gives
generally rise to $O(1/L)$ corrections. However, when studying the
effects of its perturbation in the $\kappa\to 0$ limit keeping $\kappa
L^{7/8}$ fixed, the leading scaling corrections are those arising from
the analytic expansion~\cite{CPV-14} of the scaling field $u_\kappa$
associated with the defect parameter $\kappa$. In particular at the
critical point $g=g_c$, taking into account the parity property of the
defect term, we expect
\begin{equation}
  u_\kappa\approx \kappa + c \, \kappa^3 + \ldots \,.
  \label{ukappa}
\end{equation}
Since the FSS limit is actually obtained by keeping the product
$u_\kappa L^{y_\kappa}$ fixed, involving the analytic scaling
field~\cite{CPV-14,RV-21}, the third-order correction in
Eq.~\eqref{ukappa} gives rise to $O(L^{-2y_\kappa})$ scaling
corrections, with $2y_\kappa = 7/4$, which decay more slowly than
those arising from the leading irrelevant operators of the
two-dimensional Ising universality class.

\subsection{Critical crossover of the defect fidelity}
\label{fssfidel}

A discussion on the FSS behavior of the ground-state fidelity
associated with homogeneous variations of the system
Hamiltonian, within the critical region (around $g=g_c$),
can be found in Refs.~\cite{RV-18, RV-21}.
Extending these scaling arguments to the case of localized variations,
we arrive at a scaling hypothesis for the critical nonanalytic part
at the CQT and around $\kappa=0$, i.e.
\begin{equation}
  A(L,g,\kappa,\delta \kappa)_{\rm sing} \approx
  {\cal A}(W,K,\delta K) \,,
  \quad \delta K \equiv \delta \kappa \, L^{y_\kappa}\,.
  \label{fisca}
\end{equation}
The behavior of its susceptibility $A_2$ is then obtained from
Eq.~\eqref{fisca}, by expanding ${\cal A}$ in powers of $\delta K$,
and matching it with Eq.~\eqref{expfide}:
\begin{equation}
  A_2(L,g,\kappa)_{\rm sing}
  \approx \left( {\delta K\over \delta \kappa}\right)^2 \!
          {\cal A}_2(W,K)\, \approx \, L^{2y_\kappa} {\cal A}_2(W,K)\,.
  \label{chifscal}
\end{equation}
The above asymptotic FSS behaviors are expected to be approached with
$O(L^{-2y_\kappa})$ corrections, see Sec.~\ref{scalco}.

Note that for finite values of $\kappa>0$, i.e. keeping $\kappa>0$
fixed in the large-size limit, the above FSS behavior is not expected
to hold anymore.  Indeed, since the asymptotic FSS behavior does not
change for any $\kappa>0$, we simply expect that the variation of the
fidelity is much smoother, due to the fact that the ground states for
$\kappa>0$ and $\kappa+\delta\kappa$ are expected to differ only in a
finite region around the defect. Therefore, its susceptibility is not
expected to diverge with increasing the lattice size, i.e.
\begin{equation}
A_2(L,g,\kappa>0) = O(1) \,.
  \label{chianormal}
\end{equation}

The above behaviors of the fidelity and its susceptibility will be
confirmed by numerical computations.

\section{Critical crossover driven by boundary defects}
\label{fssboufield}

We now discuss the case of Ising systems with boundaries, as such as
that defined in Eq.~\eqref{Isc} with OBC, in the presence of parallel
boundary fields as those in Eq.~\eqref{bdefbou}, which also give
rise to a relevant perturbation. Changing $\zeta$ from $0$ to
$\infty$ moves the system from the symmetric OBC to the PFBC
that violate the global ${\mathbb Z}_2$ symmetry.  The main difference
with the case of defects within rings is that the RG dimension of
boundary fields $\varphi_b$ differs from that of fields in the
bulk. Indeed, within the two-dimensional universality class, its
scaling dimension turns out to be $y_b = 1/2$~\cite{AL-91,CL-91,CZ-94},
instead of the bulk value $y_\varphi=1/8$.
Thus, using the same formula~\eqref{ykappa}, after
replacing $y_\varphi$ with $y_b$, we obtain
\begin{equation}
  y_\zeta= z - y_b = 1/2\,.
  \label{yzeta}
\end{equation}
This implies that the corresponding scaling variable is given
by~\cite{CPV-15}
\begin{equation}
  Z = \zeta\, L^{1/2}\,.
  \label{zdef}
\end{equation}
The critical crossover driven by the boundary fields is expected to
hold in the large-$L$ limit keeping $Z$ fixed, thus for $\zeta\sim L^{-1/2}$.

The critical crossover of the gap is described by the scaling
equation~\cite{CPV-15}
\begin{equation}
 \Delta(L,g,\zeta) \approx L^{-z} {\cal D}(W,Z)\,.
\label{gapo}
\end{equation}
On the other hand, its finite-$\zeta$ behavior scales as
\begin{equation}
  \Delta(L,g,\zeta>0) \approx L^{-z} {\cal D}_{\rm pfbc}(W)
  \label{infkbeho}
\end{equation}
independently of $\zeta>0$, where ${\cal D}_{\rm pfbc}(W)$ is the FSS
function of the gap with PFBC.  We again expect that
\begin{equation}
{\cal D}(W,Z\to\infty) = {\cal D}_{\rm pfbc}(W)\,.
\label{dwzinf}
\end{equation}
Therefore, systems with boundary defects of finite strength
$\zeta>0$ develop the same FSS of those with PFBC, independently of
the actual value of $\zeta$.

The scaling behavior of Eqs.~\eqref{gapo} and~\eqref{infkbeho}
has been analytically shown to hold in Ref.~\cite{CPV-15}. In particular,
at the critical point $g=g_c$, the critical crossover interpolates
between the value ${\cal D}(W,Z=0) = \pi$ [corresponding to the
  amplitude of the gap for systems with OBC, cf. Eq.~\eqref{deltafss}]
to ${\cal D}(W,Z=\infty) = 4\pi$ (corresponding to the amplitude of
the gap for systems with PFBC).

The magnetization and correlation functions are expected to behave as
\begin{eqnarray}
  M_x(x,L,g,\zeta) &\approx& L^{-y_\varphi} {\cal
    M}_x(X,W,Z)\,,
  \label{mxfsso}\\
  G(x,y,L,g,\zeta) &\approx& L^{-2y_\varphi}
  {\cal G}(X,Y,W,Z)\,,
  \label{gxyfsso}
\end{eqnarray}
where $X=x/L$ and $Y=y/L$.  Analogous FSS equations are obtained for
the other quantities defined in Sec.~\ref{obser}. For example we may
consider the most natural definition of second-moment correlation
(\ref{chixidef}), taking $\xi \equiv \xi_x$ with $x$ located at the
center of the chain, and obtain
\begin{equation}
  R_\xi(L,g,\zeta) \equiv \xi/L \approx {\cal R}_\xi(W,Z)\,.
  \label{rxiscao}
\end{equation}

We can also derive scaling formulas for the ground-state fidelity
associated with the boundary fields, analogous to those for the Ising
ring with one defect, by replacing $K$ with $Z$, and $\delta K$ with
$\delta Z$:
\begin{eqnarray}
  A(L,g,\zeta,\delta \zeta)_{\rm sing} &\approx& {\cal A}(W,Z,\delta
  Z) \,, \quad \delta Z \equiv \delta \zeta \, L^{y_\zeta}\,,\qquad
  \label{fisca2}\\
  A_2(L,g,\zeta) &\approx& L^{2y_\zeta} {\cal A}_2(W,Z)\,.
  \label{cqtchil2}
\end{eqnarray}
Note that even boundary defects give rise to a power-law growth of
$A_2$ when increasing $L$, but this is significantly slower than the
case of bulk defects, indeed $A_2 \sim L$ since $y_\zeta=1/2$.  Again
for finite fixed $\zeta$ we expect $A_2(L,g,\zeta) = O(1)$.

The power-law approach to the above asymptotic FSS behaviors can be
inferred by the analysis reported in Sec.~\ref{scalco}.  The leading
scaling corrections are generally expected to be $O(L^{-1})$, arising
from the presence of the boundaries and the analytic expansion of the
scaling field associated with $\zeta$ (since $2 y_\zeta=1$). A slower
approach should still characterize the observables involving
second-moment correlation length, as $L^{-3/4}$, due to background
contributions.

We finally note  that Ising chains with OBC and in the presence of
symmetry-breaking defects in the bulk (i.e. far from the boundaries),
such as those described by the Hamiltonian term (\ref{dxdef}), are
expected to develop a critical crossover behavior driven by the defect
as well, similar to that of quantum Ising rings, already discussed in
Sec.~\ref{critcross}. While the scaling behavior is still controlled
by the RG dimension $y_\kappa=7/8$, the corresponding scaling
functions are expected to differ, because they interpolate between a
system with OBC (when $\kappa=0$) and the $\kappa\to\infty$ limit
consisting of a system with two subsystems of size $L_1$ and $L_2$
(where $L_1,\,L_2$ are the distances of the defect from the
boundaries) having mixed boundary conditions: OBC on one side and
fixed boundary conditions on the other one (corresponding to the
position of the defect).

\section{Numerical results}
\label{numres}

To support the scaling behaviors put forward in the previous sections,
we now present some numerical results, obtained by exact diagonalization
(up to $L=20$ sites) and by density-matrix renormalization group (DMRG)
for larger systems (up to $L=40$ --- see App.~\ref{dmrg} for details
on the implementation for systems with PBC).

\subsection{Critical defect crossover in Ising rings} 
\label{isringdef}

\begin{figure}[hbtp]
    \centering \includegraphics[width=0.95\columnwidth,
      clip]{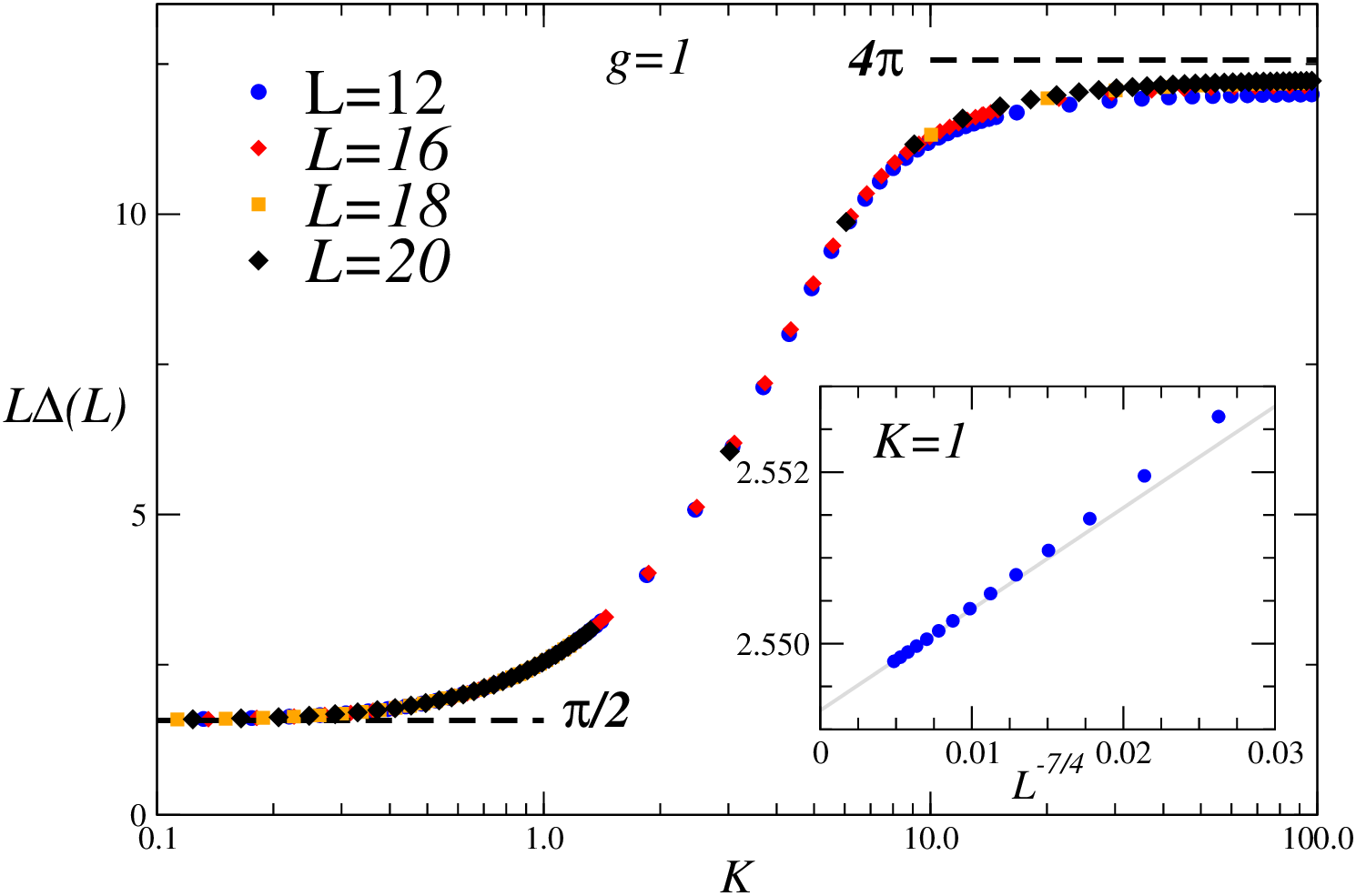}
    \caption{Critical crossover of the gap $\Delta(L,g,\kappa)$ at the
      critical point $g_c=1$. The collapse of the data of $L\,\Delta$
      versus $K=\kappa L^{y_\kappa}$ supports the scaling
      behavior reported in Eq.~(\ref{gapfss}). The dashed lines
      indicate the limiting cases: ${\cal D}(0,K\to 0) = \pi/2$ and
      ${\cal D}(0,K\to\infty) = 4\pi$, which can be obtained from
      Eq.~(\ref{deltafss}), corresponding to PBC and PFBC.  The inset
      shows that the scaling corrections at $K=1$ are consistent with
      the expected $O(L^{-7/4})$ suppression, see Sec.~\ref{scalco}
      (the grey line is drawn to guide the eye).}
    \label{delta_crossover}
\end{figure}

\begin{figure}[hbtp]
  \centering \includegraphics[width=0.95\columnwidth,
    clip]{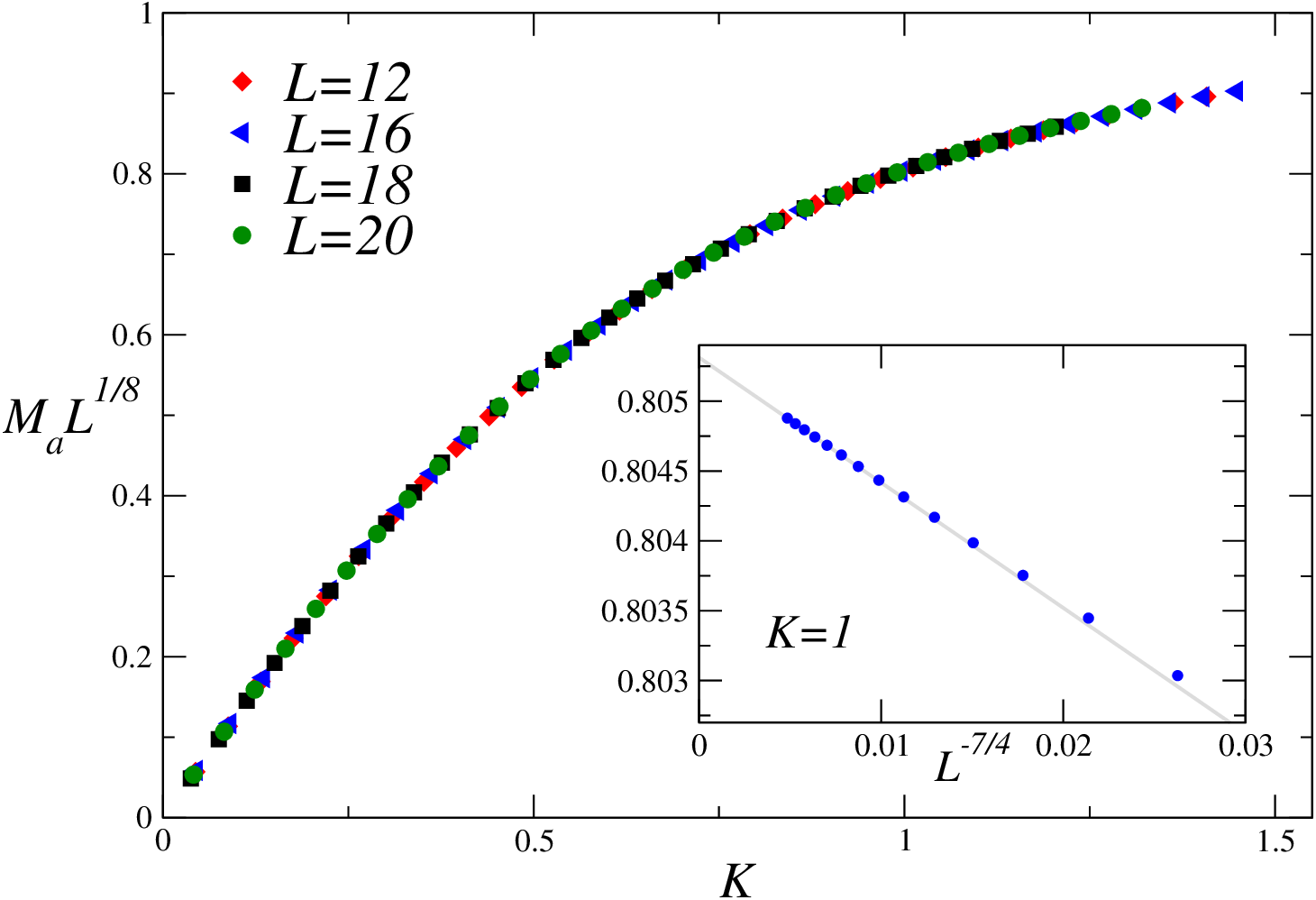}
  \caption{Critical crossover of the average magnetization $M_a$,
    defined in Eq.~(\ref{magdef}), at the critical point $g_c=1$.  The
    figure shows the rescaled data $L^{1/8} M_a$ versus $K=\kappa
    L^{7/8}$. The collapse of the data for various lattice sizes
    along a universal FSS curve supports the scaling in Eq.~\eqref{mfss}.
    The inset shows that the scaling corrections at a fixed value $K=1$
    are consistent with the expected $O(L^{-7/4})$ behavior, see Sec.~\ref{scalco}
    (the gray line is drawn to guide the eye).}
    \label{magnetization_k}
\end{figure}

\begin{figure}[hbtp]
    \centering \includegraphics[width=0.95\columnwidth,
      clip]{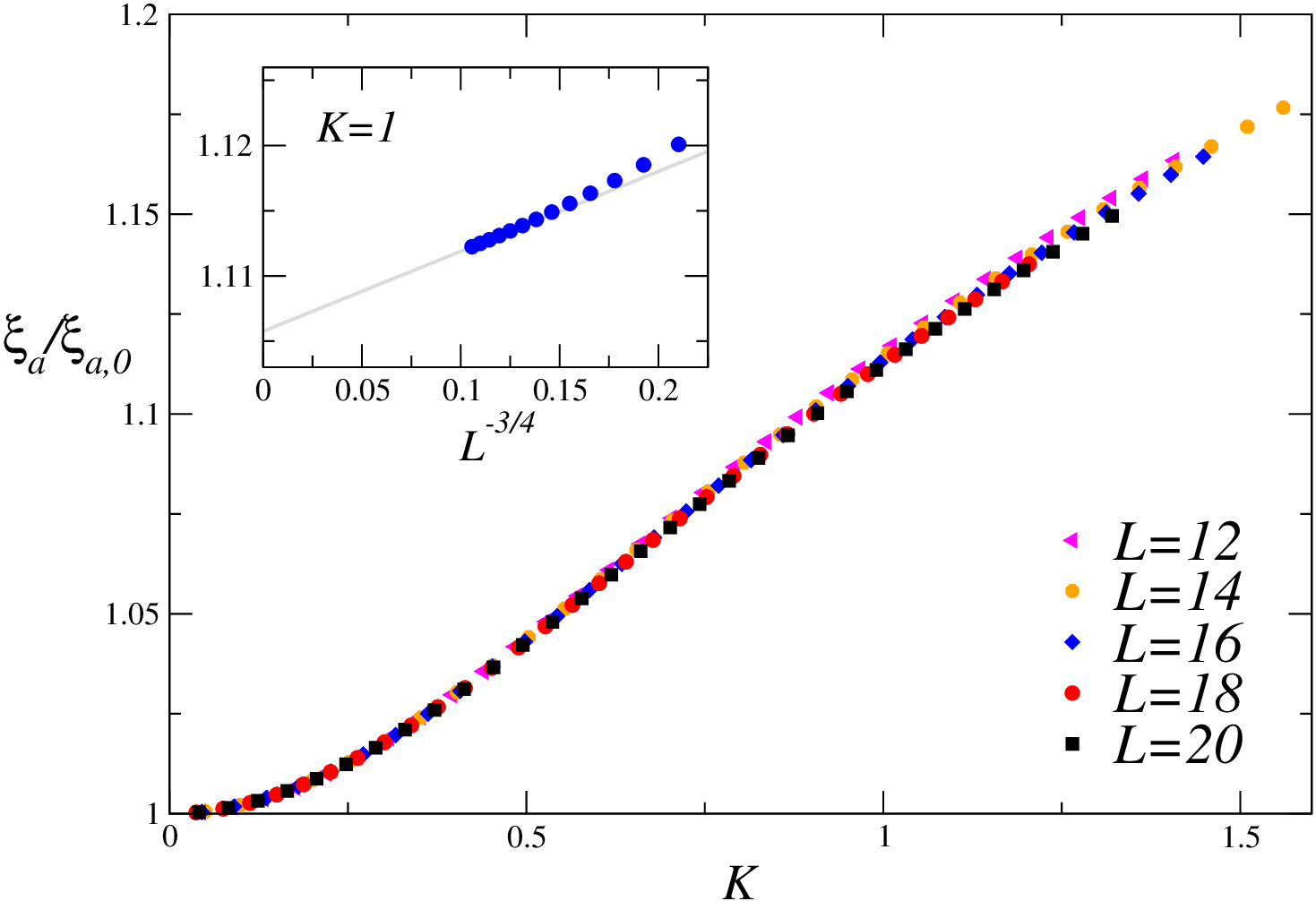}
    \caption{ The correlation length $\xi_a$, defined in
      Eq.~(\ref{avedef}), along the critical defect crossover at the
      critical point $g_c=1$. We plot the ratio $\xi_a/\xi_{a,0}$
      versus $K=\kappa L^{7/8}$, where $\xi_{a,0}$ is the correlation
      length at $\kappa=0$, see also Eq.~(\ref{xiaxia0}).  The data
      approach an asymptotic scaling function, in agreement
      with the scaling equation~\eqref{Rfss}.  The inset shows that
      the scaling corrections at a fixed value $K=1$ are consistent
      with the asymptotic power-law $O(L^{-3/4})$ suppression, see
      Sec.~\ref{scalco}.  (the gray line is drawn to guide the eye).}
    \label{scaling_rxi}
\end{figure}

We first report results supporting the critical crossover behavior
driven by the symmetry-breaking defect~\eqref{dxdef} in quantum Ising
rings, as discussed in Sec.~\ref{critcross}.  These are obtained around
the critical point $g_c=1$; actually most of them are exactly at $g=g_c$.

The energy difference between the two lowest levels (i.e., the gap)
at the critical point is reported in Fig.~\ref{delta_crossover}.
The behavior of the various curves for different sizes $L$ matches
the scaling ansatz~\eqref{gapfss}. In particular, data for $L \, \Delta(L)$
as a function of $K=\kappa L^{y_\kappa}$ range from
$\pi/2$ to $4\pi$, consistently with the crossover from PBC to PFBC.
In fact, the amplitude $C_\Delta$ of the leading behavior $\Delta\approx C_\Delta/L$
at the critical point goes from $C_\Delta=\pi/2$, for systems without
boundaries corresponding to $\kappa=0$, to $C_\Delta=4\pi$, in the
$K\to\infty$ limit of PFBC corresponding to finite $\kappa$.

In Fig.~\ref{magnetization_k} we report some results for the averaged
magnetization defined in Eq.~\eqref{magdef} at $g=g_c$. While the
bare data points behave differently for various system sizes, a nice
data collapse when plotting $L^{1/8} M_a$ versus $K=\kappa L^{7/8}$ is observed,
thus supporting the FSS ansatz~\eqref{mfss}.
Leading scaling corrections are $O(L^{-7/4})$, coming from the
analytical expansion~\eqref{ukappa} of the defect scaling field.

Further evidence of the critical crossover is provided by
a numerical analysis of the ratio
\begin{equation}
  {\xi_a\over \xi_{a,0}}\equiv
    {\xi_a(L,g=g_c,\kappa)\over \xi_{a}(L,g=g_c,\kappa=0)}
    = {R_\xi(L,g=g_c,\kappa)
    \over R_\xi(L,g=g_c,\kappa=0)}\,,
  \label{xiaxia0}
\end{equation}
presented in Fig.~\ref{scaling_rxi}.
The data appear to approach an asymptotic scaling function, in
agreement with the scaling equation~\eqref{Rfss}.  In this case, the
leading scaling corrections to $R_\xi$ are $O(L^{-3/4})$, see also the
inset of Fig.~\ref{scaling_rxi}, coming from the analytical background
at the CQT (see the discussion in Sec.~\ref{scalco}).

The above results definitely confirm the scaling predictions for the
critical crossover behavior driven by the defects, moving away from
the translation-invariant and ${\mathbb Z}_2$-symmetric FSS of the
critical Ising ring~\eqref{Iring}.

\subsubsection{The defect ground-state fidelity}
\label{defgrfide}

\begin{figure}[hbtp]
    \centering \includegraphics[width=0.95\columnwidth,
      clip]{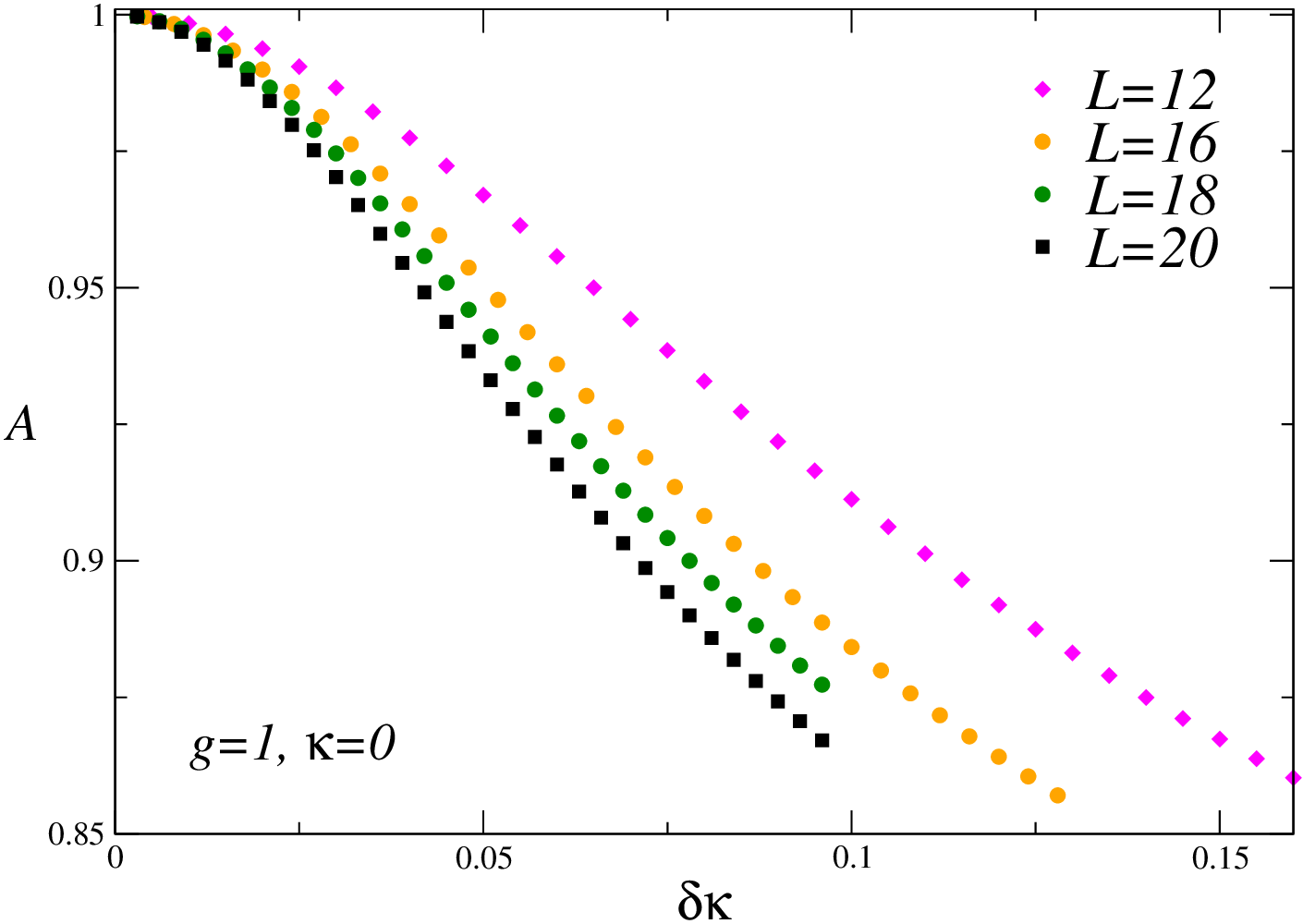}
    \includegraphics[width=0.95\columnwidth, clip]{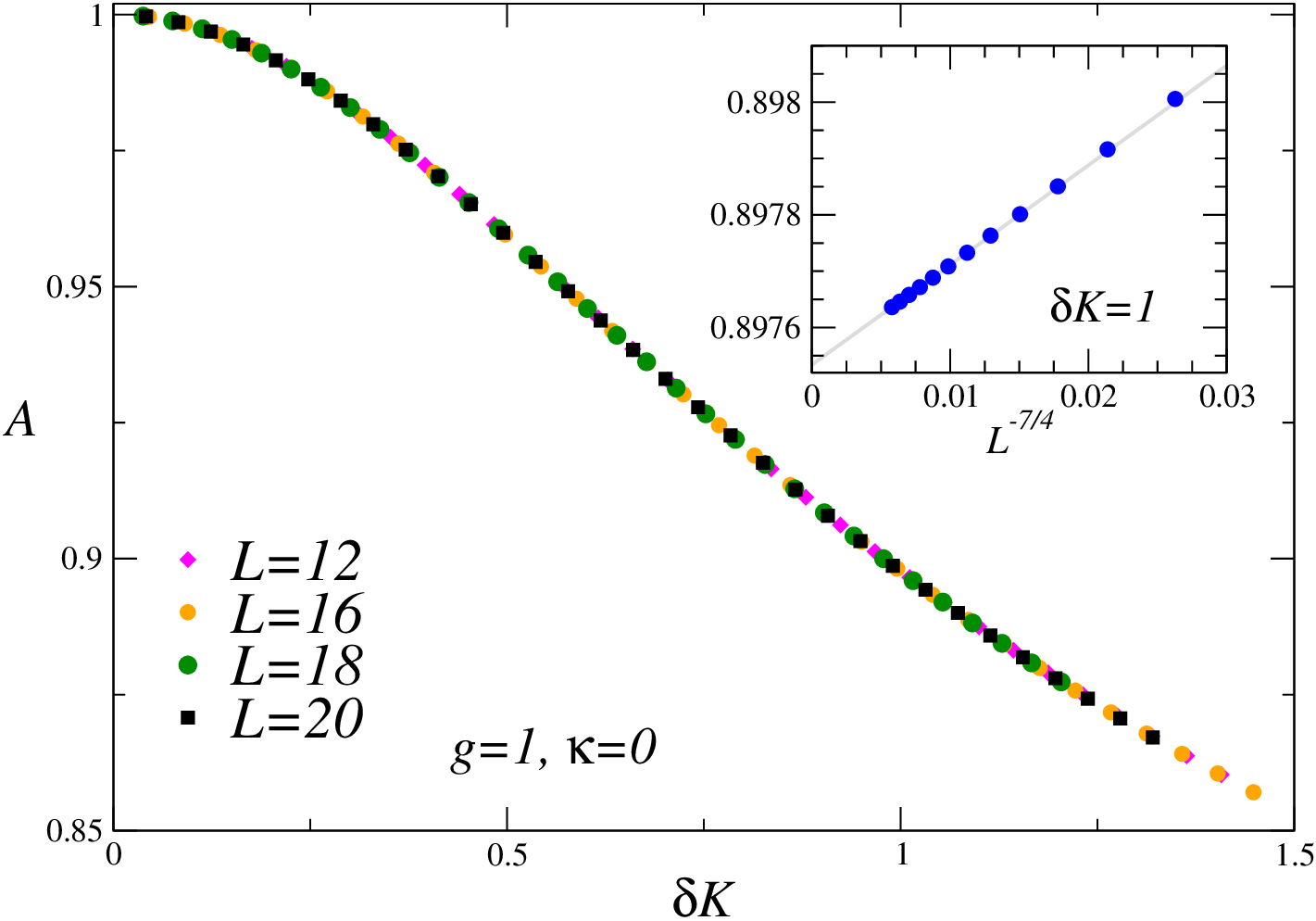}
    \caption{The ground-state fidelity (\ref{fiddef}) associated with
      the defect, at the critical point $g_c=1$. The top figure
      reports the plain data for $\kappa=0$ as a function of
      $\delta\kappa$, showing that they tend to get suppressed at
      smaller and smaller values of $\delta\kappa$ when increasing the
      size of the system.  The bottom figure shows them versus $\delta
      K=\delta\kappa L^{7/8}$. The clear collapse of the curves
      definitely supports the scaling behavior (\ref{fisca}). The
      inset of the bottom figure shows show that the scaling
      corrections at $\delta K=1$ are consistent with the expected
      $O(L^{-7/4})$ decay, see Sec.~\ref{scalco}.}
    \label{fidelity_susc_fig_k0}
\end{figure}

\begin{figure}[hbpt]
    \centering
    \includegraphics[width=0.95\columnwidth, clip]{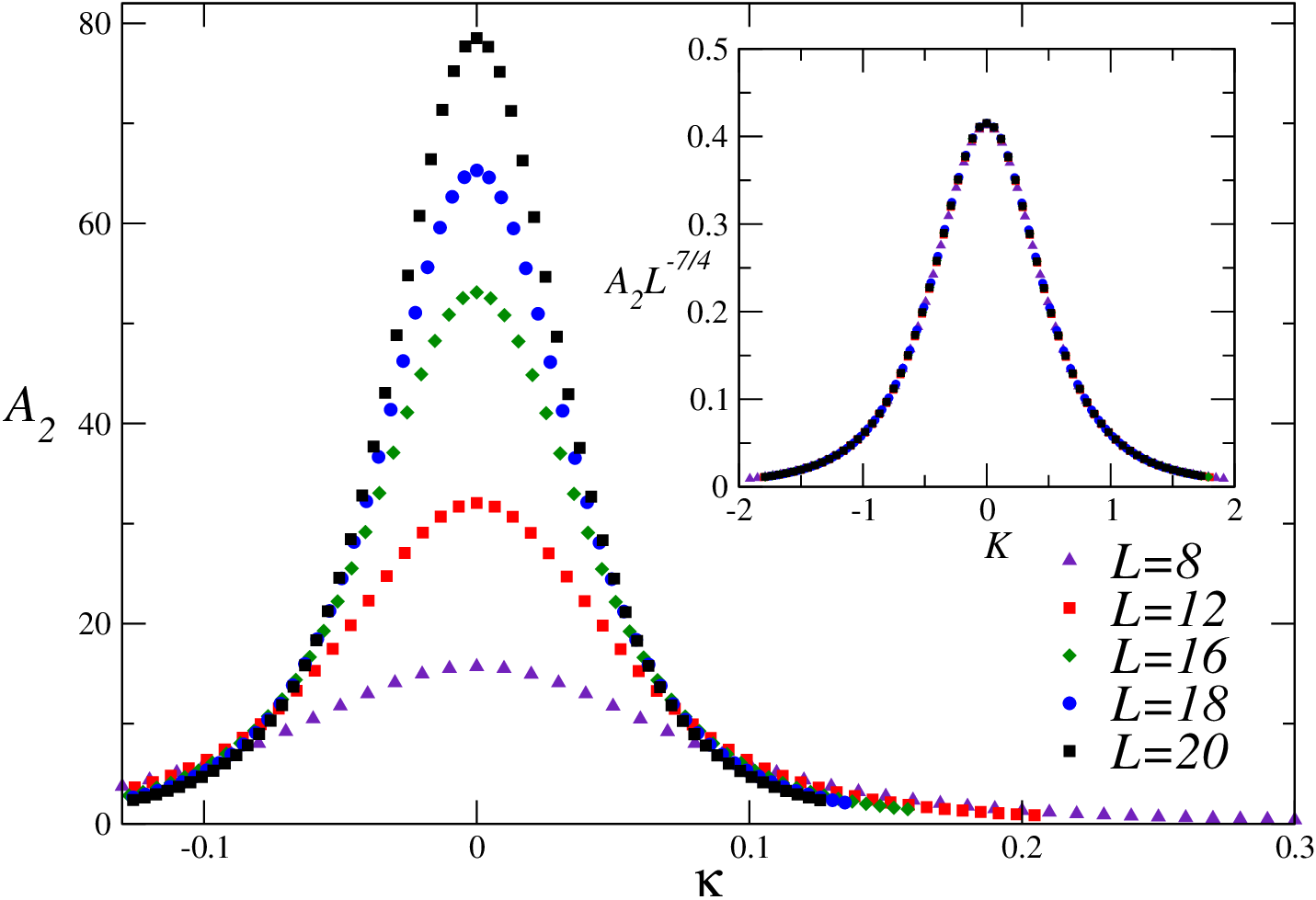}
    \caption{The fidelity susceptibility $A_2$,
      cf. Eq.~\eqref{expfide}, associated with one defect within Ising
      rings, at the critical point $g_c=1$ and around $\kappa=0$.
      Note that, by symmetry, $A_2$ is invariant under $\kappa\to
      -\kappa$.  The plain data already suggest that $A_2$ around
      $\kappa=0$ diverges with increasing $L$. The inset shows a plot
      of $L^{-2y_\kappa} A_2$ versus $K=\kappa L^{2y_k}$ with
      $y_\kappa=7/8$, which represents a robust evidence in favor of
      the scaling behavior (\ref{chifscal}), and in particular the
      $L^{7/4}$ divergence with increasing $L$.  }
    \label{susc_k}
\end{figure}

Interesting features of the critical crossover driven by defects
emerge when looking at the ground-state fidelity
[cf., Eq.~\eqref{fiddef}].
As discussed in Sec.~\ref{fssfidel}, we
expect that the fidelity, and in particular the associated
susceptibility (\ref{a2def}), exhibits qualitatively different
behaviors around $\kappa=0$ and for any finite and fixed $\kappa>0$.

In Fig.~\ref{fidelity_susc_fig_k0} we report some results for the
fidelity $A$ at the critical point and for $\kappa=0$, as a function
of $\delta\kappa$.  The top panel shows the fidelity as a function
of $\delta\kappa$, which appear suppressed at smaller and smaller
values of $\delta\kappa$, when increasing the system size.
Plotting the same data versus $\delta K=\delta\kappa L^{y_\kappa}$,
with $y_\kappa=7/8$, a nice collapse toward an asymptotic curve emerges,
thus supporting the FSS behavior~\eqref{fisca} (bottom panel).
Corrections to the asymptotic FSS are suppressed consistently
with the power law $L^{-7/4}$, as expected from the analysis of Sec.~\ref{scalco}.

\begin{figure}[hbtp]
    \centering
    \includegraphics[width=0.95\columnwidth, clip]{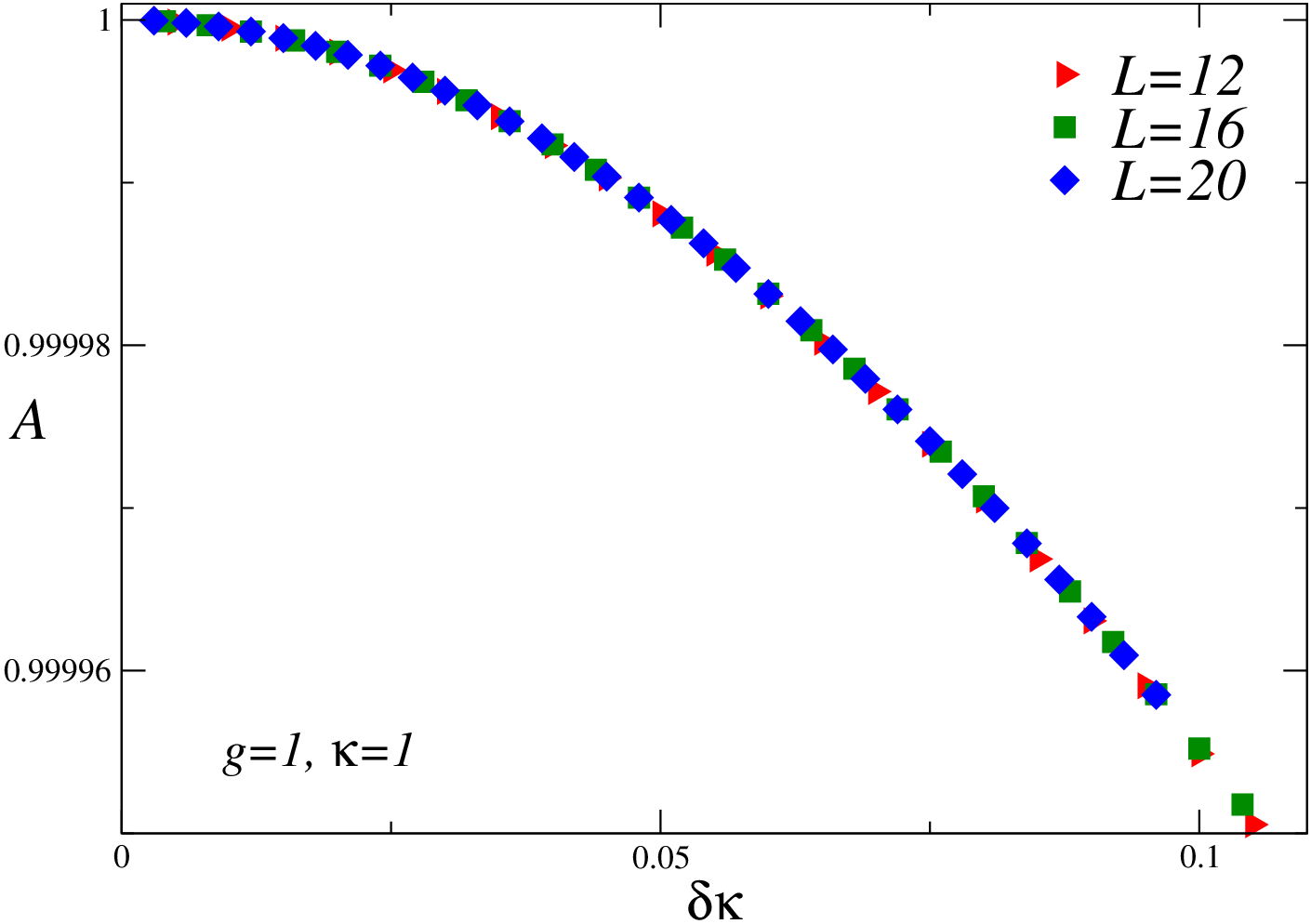}
    \caption{The ground-state fidelity at the critical point, for
      $\kappa=1$ versus $\delta \kappa$. With increasing system
      size, the curves become rapidly independent of $L$,
      thus implying that the corresponding fidelity
      susceptibility $A_2$ remains finite with increasing $L$.}
    \label{fidelity_1k_fig}
\end{figure}

Figure~\ref{susc_k} shows results for the fidelity susceptibility
$A_2$ at the critical point $g=g_c$~\cite{Note1}.  They confirm the
scaling Eq.~(\ref{chifscal}), and in particular that in the critical
crossover regime $A_2$ diverges as $L^{2y_\kappa}=L^{7/4}$ with
increasing $L$. This shows that, at the critical point and around
$\kappa=0$, the impact of the defect~\eqref{dxdef} on the system
ground state is quite strong.

Finally we have also done the analogous computations for $\kappa=1$,
therefore relatively far from the critical crossover region where
$\kappa\sim L^{-7/8}$ involved in the critical defect crossover around
$\kappa=0$.  As shown in Fig.~\ref{fidelity_1k_fig}, the ground-state
fidelity for $\kappa=1$ and $\delta\kappa>0$ becomes rapidly
independent of $L$ with increasing the system size. This
implies that the corresponding fidelity susceptibility $A_2$ remains
finite with increasing $L$, as predicted by
Eq.~\eqref{chianormal}. This behavior turns out very different from
the anomalous $L^{2y_k}$ divergence characterizing the critical
crossover behavior around $\kappa=0$.

\subsection{Critical crossover arising from boundary defects}
\label{defgrfidebf}

\begin{figure}[hbpt]
    \centering
    \includegraphics[width=0.95\columnwidth, clip]{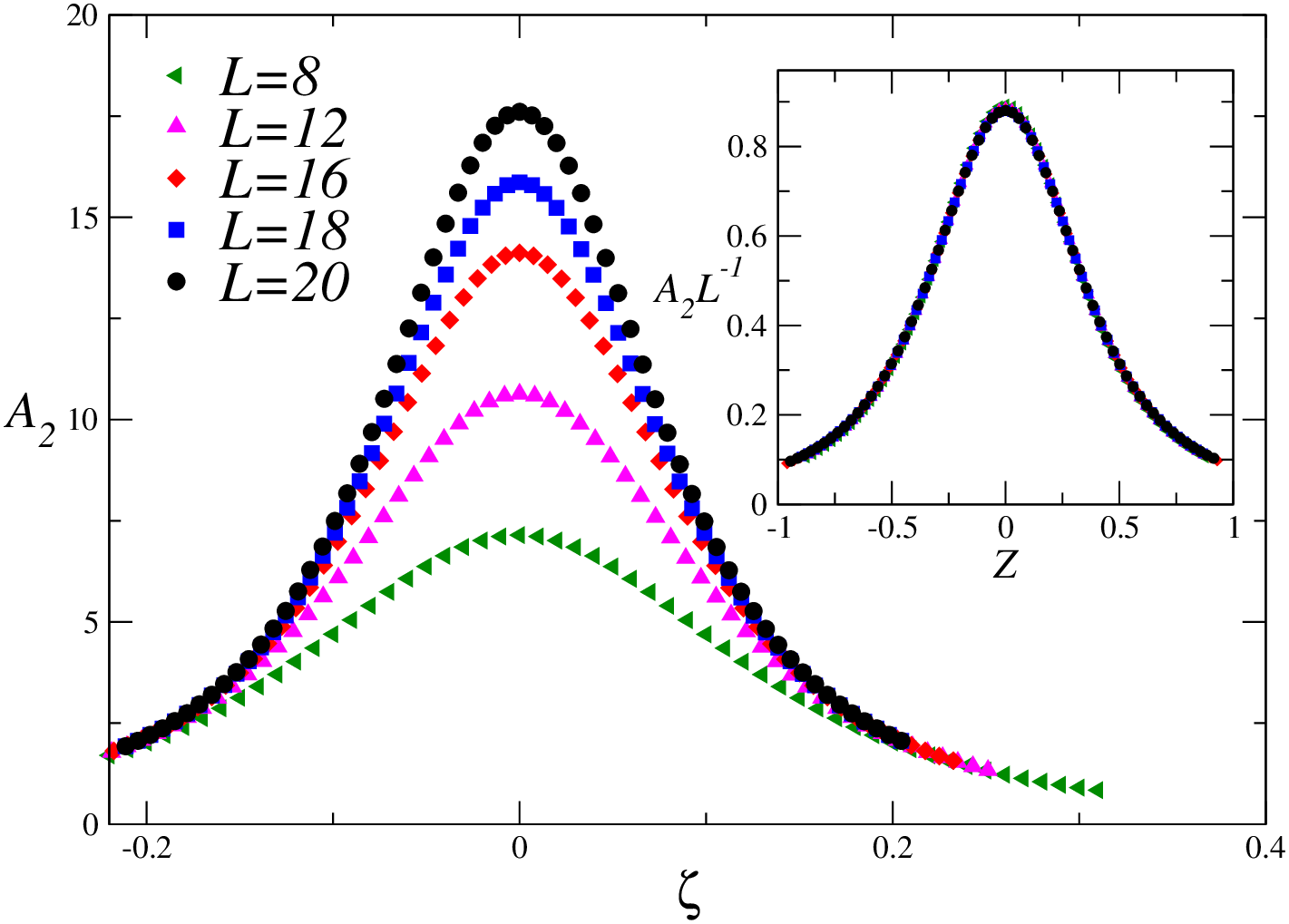}
    \caption{The fidelity susceptibility $A_2$, associated with the
      boundary defects within Ising chains with OBC, at the critical
      point $g_c=1$ and around $\zeta=0$. The inset shows a plot of
      $L^{-2 y_\zeta}A_2$ versus $Z=\zeta L^{y_\zeta}$ with
      $y_\zeta=1/2$.  The data clearly support the RG prediction that
      $A_2$ diverges as $L^{2y_\zeta}$, within the
      critical crossovger region, keeping $Z$ fixed.}
    \label{susc_k_bf}
\end{figure}

We now consider boundary defects as those arising from boundary fields
in Ising chains, cf. Eq.~\eqref{bdefbou}.
Some results for the critical crossover of the gap have
been alredy reported in Ref.~\cite{CPV-15}.  Here we supplement them
with results for the ground-state fidelity measuring the overlap of
the ground states for different values of the boundary-field parameter
$\zeta$.

\begin{figure}[hbtp]
    \centering
    \includegraphics[width=0.95\columnwidth, clip]{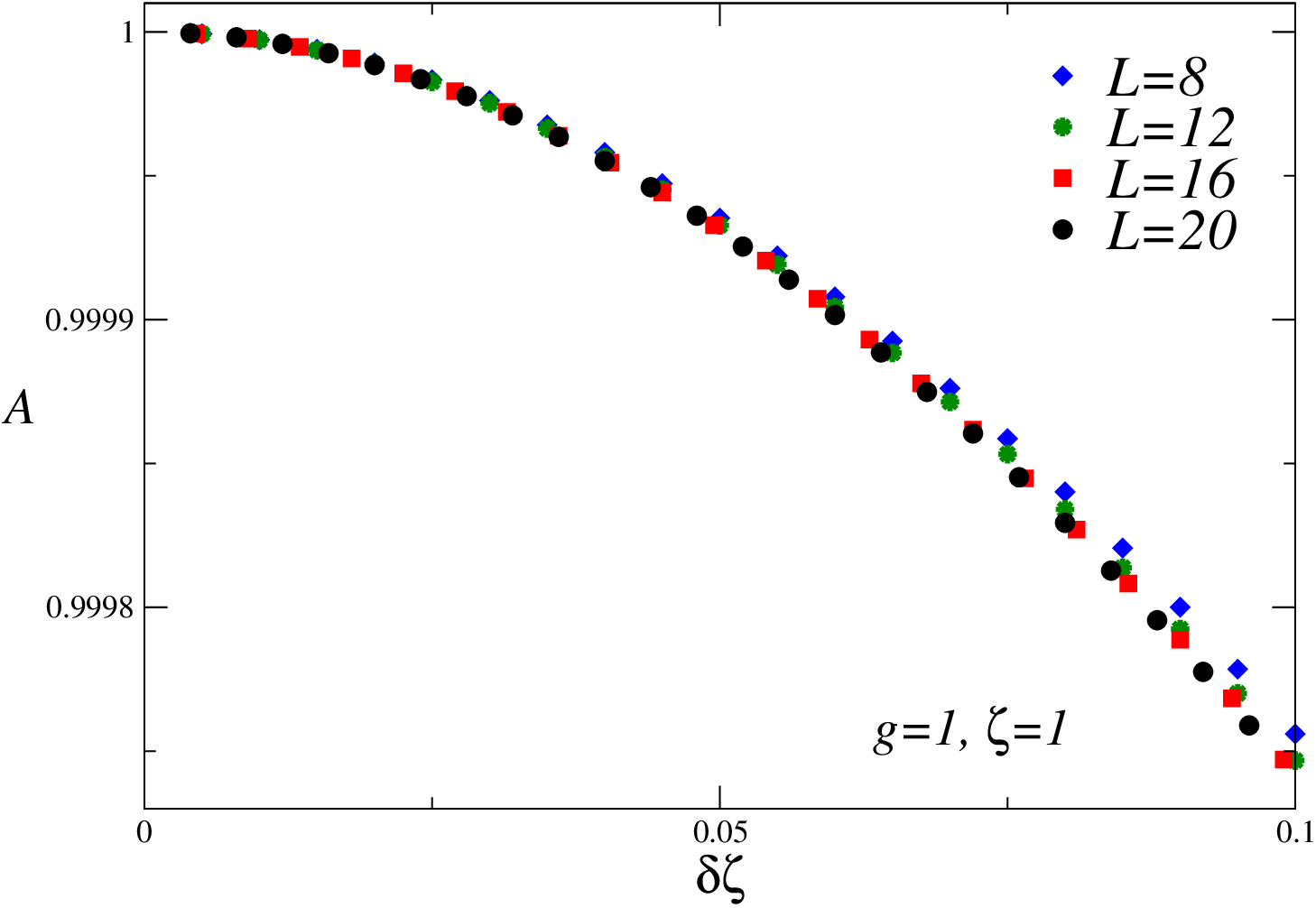}
    \caption{ The ground-state fidelity at the critical point, for
      $\zeta=1$ versus $\delta \zeta$. With increasing the system size,
      the curves become rapidly independent of $L$, implying that the
      corresponding fidelity susceptibility $A_2$ remains finite
      with increasing $L$.  This behavior is very different
      from the $L^{2y_\zeta}$ divergence characterizing the critical crossover
      behavior around $\zeta=0$, see Fig.~\ref{susc_k_bf}.}
    \label{fidelity_1kbf_fig}
\end{figure}

As discussed in Sec.~\ref{fssboufield}, we expect a critical crossover
scenario analogous to that found for Ising rings, with the main
difference that the fidelity susceptibility is expected to diverge as
$L^{2y_\zeta}$, thus as $L$, with increasing the size, at criticality
and around $\zeta=0$.  The results in Fig.~\ref{susc_k_bf} nicely
confirm these predictions.
We also explicitly checked that corrections to the asymptotic FSS
are suppressed as $L^{-1}$, as expected from the analysis of Sec.~\ref{scalco}
(data not shown).
On the other hand, like for bulk defects, the fidelity susceptibility
for finite $\zeta>0$ converges to a constant. This is demonstrated
by the data for the fidelity at $\zeta=1$ as a function of $\delta\zeta$,
shown in Fig.~\ref{fidelity_1kbf_fig}, which appear to rapidly converge
to a function of $\delta \zeta$ independent of $L$.

\subsection{FSS keeping the defect strength fixed} 
\label{fssfiniteka}

\begin{figure}
  \centering
    \includegraphics[width=0.95\columnwidth,
      clip]{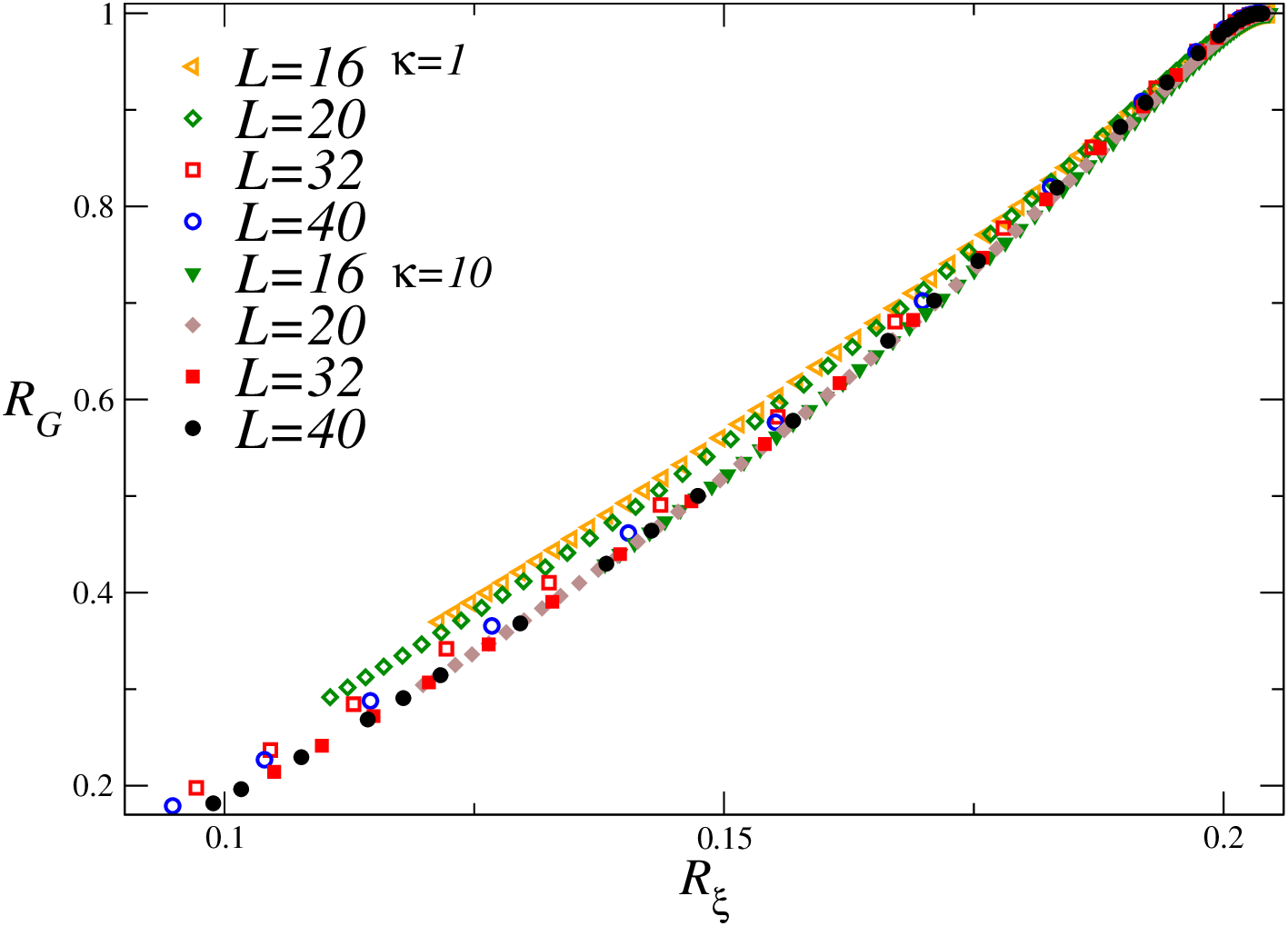}
    \caption{Plots of $R_{G}$ versus $R_{\xi}$, see text, for
      $\kappa=1, 10$. Data for $\kappa=1$ and $\kappa=10$ appear to
      approach the same asymptotic curve. Data for $L=32, 40$ are
      obtained by means of DMRG.}
    \label{rgrxikappa}
\end{figure}

\begin{figure}
  \centering
  \includegraphics[width=0.95\columnwidth,
      clip]{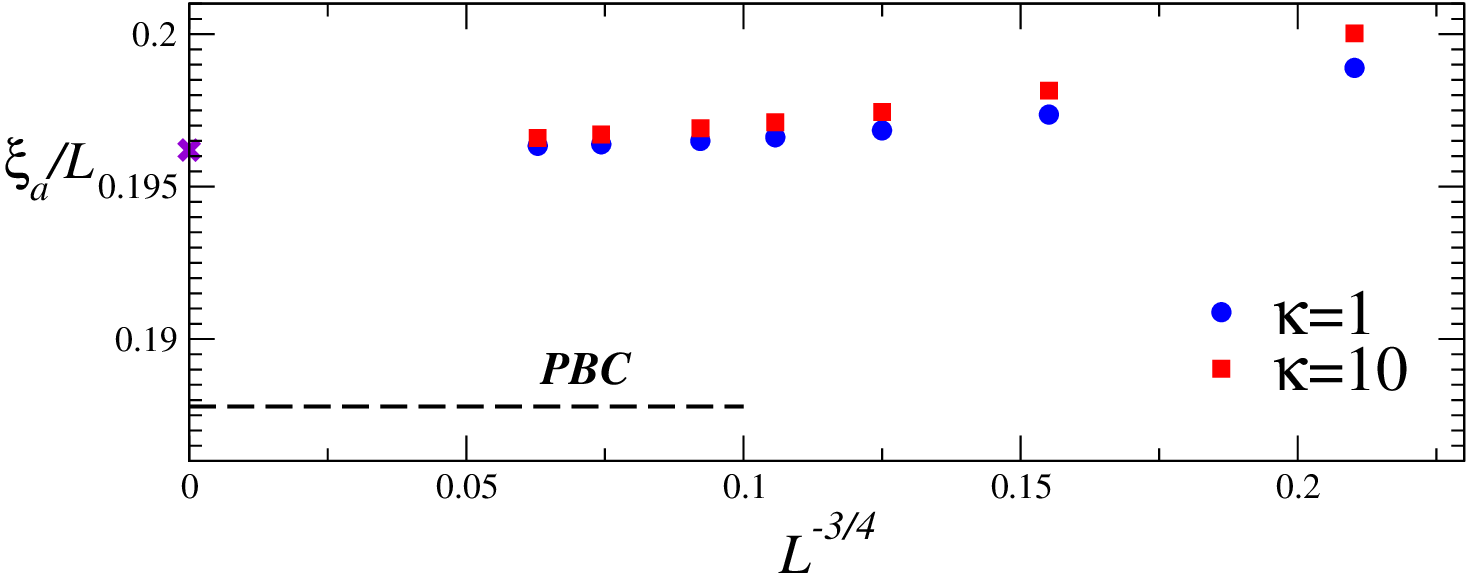}
    \caption{Data for $R_{\xi a}$ versus $L^{-3/4}$ at the critical
      point and for $\kappa=1,10$, compared with the value for
      $\kappa=0$ given by the exact result $R_{\xi
        a}^\star=0.187789...$ (indicated by the dashed line), easily
      obtained from the critical two-point
      function~\cite{DMS-book}. The large-$L$ extrapolations for
      $\kappa=1,10$ are compatible with each other, and approximately
      equal to $R_{\xi a}^\star\approx0.1962$ (violet cross in the
      plot), differing from the value at $\kappa=0$. They support
      the fact that the FSS at finite fixed $\kappa>0$ is expected
      to be independent of $\kappa$.}
    \label{rxi_extrapolation}
\end{figure}

\begin{figure}[hbtp]
    \centering \includegraphics[width=0.95\columnwidth,
      clip]{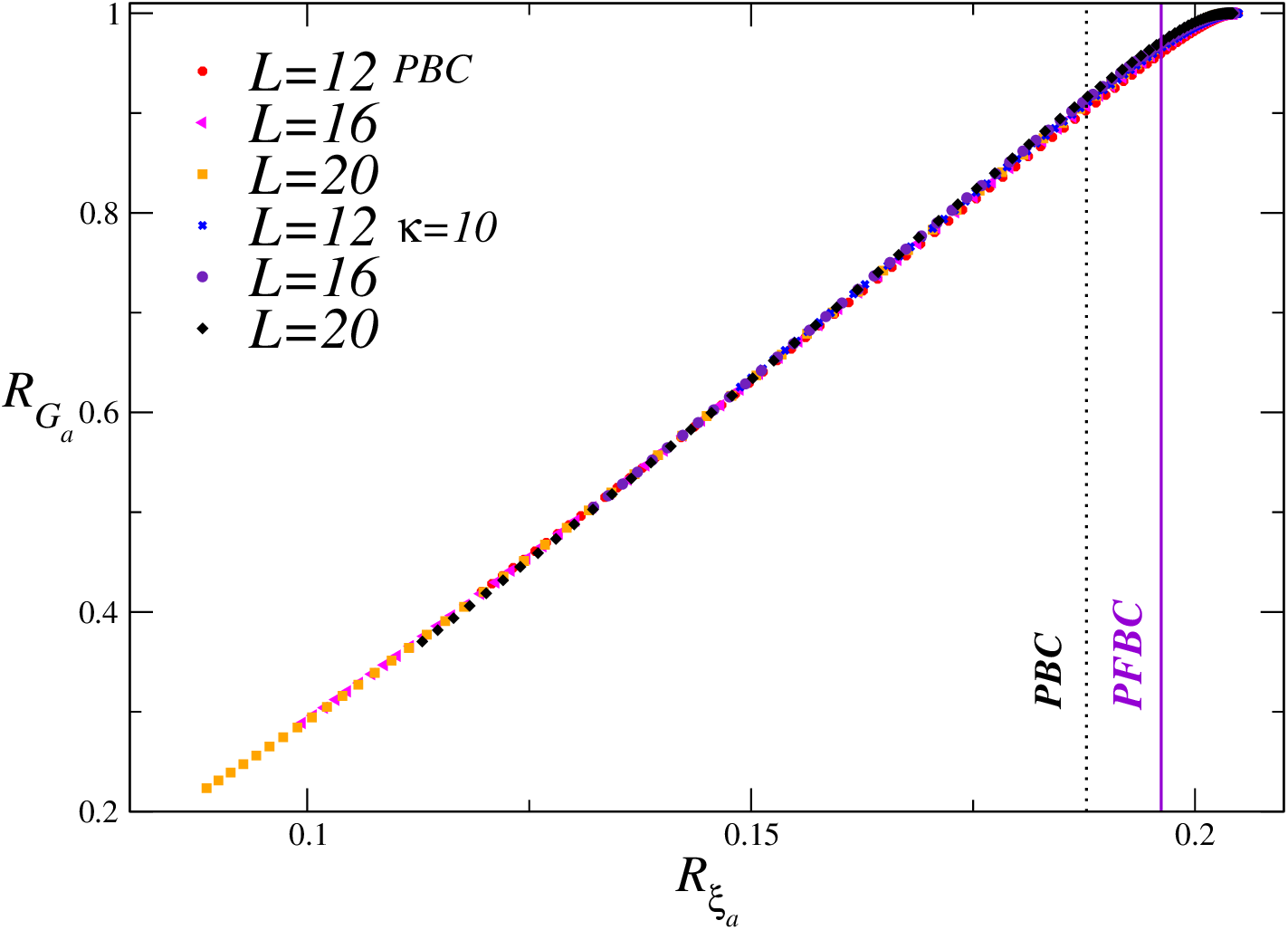}
    \caption{The averaged quantities $R_{G a}$ versus $R_{\xi a}\equiv \xi_a/L$.
      The vertical dotted-black and straight-violet lines represent,
      respectively, the $R_{\xi a}^\star$ critical values for PBC and
      PFBC, corresponding to the limiting cases of critical defect crossover
      for $\kappa=0$ and $\kappa\to\infty$.  Although their values at
      $g=g_c$, indicated by the vertical lines, differ significantly,
      see also Fig.~\ref{rxi_extrapolation}, the FSS curves of the
      averaged quantities turn out to be very similar when moving from
      $\kappa=0$ to a finite $\kappa$.}
    \label{rxi_and_U}
\end{figure}

Let us go back to the quantum Ising ring (\ref{Iring}) with one
symmetry-breaking defect (\ref{dxdef}). As a final check of the
critical crossover scenario, we provide numerical evidence that the
asymptotic FSS for $\kappa>0$ does not depend on $\kappa$, and it
corresponds to that for $\kappa\to\infty$, when the system becomes
equivalent to an Ising chain with PFBC. In other words, the FSS limit
keeping $\kappa>0$ fixed must be independent of $\kappa$, as discussed
at the end of Sec.~\ref{fssobs}, see in particular
Eq.~(\ref{Rfsskagt0}).

To avoid problems arising from possible change of normalizations of
the scaling variable $W$, we proceed as follows.  We consider the RG
invariant quantities introduced in Sec.~\ref{obser}. In particular, we
consider $R_{\xi}\equiv R_{\xi x}$, $R_{G}\equiv R_{G x}$ and $U\equiv
U_{x}$, cf. Eqs.~(\ref{rxidef}), (\ref{rgdef}) and (\ref{uxdef}), with
$x$ given by the position of the defect.  Their FSS behavior for
$\kappa>0$ must generally be $R(L,g) \approx {\cal R}(W)$
independently of $\kappa$.  Since $R_\xi$ is a monotonic function, for
$\kappa>0$, we may also write
\begin{equation}
  U(L,g,\kappa) \approx F_U(R_\xi)\,,\quad
  R_{G}(L,g,\kappa) \approx F_G(R_\xi)\,,
  \label{ugdef}
\end{equation}
where $F_U$ and $F_G$ depend on the universality class only, without
free normalizations. Numerical results in Fig.~\ref{rgrxikappa},
for $\kappa=1$ and $\kappa=10$, confirm that $R_G$ approaches
the same scaling function of $R_\xi$, independently of $\kappa$.
Analogous evidence has been found for $U$ (not shown).

As a further check, in Fig.~\ref{rxi_extrapolation} we show data for
the averaged quantity $R_{\xi a}$ defined in Eq.~(\ref{rxidef})
evaluated at $g=g_c$, for $\kappa=1$ and $\kappa=10$. They show that
the large-$L$ extrapolations using the expected $L^{-3/4}$ asymptotic
behavior, are compatible between $\kappa=1$ and $\kappa=10$, and are
definitely different from the value at $\kappa=0$.
Therefore, concerning the behavior at finite $\kappa>0$, outside the
critical crossover regime where $\kappa\sim L^{-7/8}$, the numerical
results confirm that the asymptotic FSS behavior is independent of the
defect strength $\kappa>0$, whenever they are computed keeping
$\kappa$ fixed.

Finally, we note that the average quantities $R_{G a}$ and $U_a$
[defined in Eqs.~(\ref{rgadef}) and (\ref{uxdefave})] when plotted
versus $R_{\xi a}$ [defined in Eq.~(\ref{rxidef})] show universal
curves that look very similar when moving from $\kappa=0$ to a
finite $\kappa$, i.e.  from PBC to PFBC.  This is shown, e.g., in
Fig.~\ref{rxi_and_U}, where we plot $R_{G a}$ versus $R_{\xi a}$ for
$\kappa=0$ and $\kappa=10$.  Actually, this is quite unexpected:
although the values of $R_{G a}$ and $R_{\xi a}$ at $g=g_c$ differ
significantly ($R_{\xi a}\approx 0.1878$ for $\kappa=0$ and $R_{\xi a}
\approx 0.1962$ for $\kappa=10$), the FSS curves of the averaged
quantities for $\kappa=0$ and $\kappa>0$ turn out to be very close.

\section{Conclusions}
\label{conclu}

We have investigated the effects of symmetry-breaking defects at CQTs,
arising from localized external fields coupled to the order-parameter
operator.  At CQTs the presence of isolated defects does not generally
change the bulk power-law behaviors of observables at large
scale. However, when the defects are the only source of
symmetry breaking (i.e., the original lattice system is strictly
symmetric under the global symmetry, without boundaries or with
boundaries preserving the symmetry), they drive critical crossover
behaviors entailing substantial and rapid changes of the ground-state
and low-energy properties.  The limiting cases of these critical defect
crossovers can be associated with different boundary conditions: from
boundary conditions preserving the global symmetry to ones breaking
the symmetry. Therefore, the addition of symmetry-breaking defects can
drive relevant effects in finite-size systems (or in the neighborhood
of the defect), leading to substantial changes in the finite-size
behavior of the low-energy critical modes, even in the large-size
limit within the FSS regime around the CQT.  Two different situations
must be distinguished: whether the defects are located within the bulk
of the system, or at the boundaries. Indeed, they lead to scaling
scenarios controlled by different RG exponents associated with the
universality class of the CQT.

The above scenario has been investigated within the paradigmatic
one-dimensional quantum Ising models in a transverse field, whose CQT
is related to the spontaneous breaking of a global ${\mathbb Z}_2$
symmetry.  We analyze the effects of localized defects breaking the
global ${\mathbb Z}_2$ symmetry, arising from external longitudinal
fields localized at one site of the system. We consider both bulk and
boundary defects.

Using standard RG arguments within FSS frameworks, we develop a
scaling theory to describe the critical crossover behaviors driven by
the symmetry-breaking defects. In particular, one
localized symmetry-breaking defect in critical Ising rings turns out
to develop a critical crossover between translation-invariant Ising
rings without boundaries and Ising systems with fixed and parallel
boundary conditions. We discuss the critical crossover behavior of
several observables, such as the magnetization and the correlation
function of the longitudinal spin variables.

An effective characterization of the critical defect crossover is
achieved by analyzing the ground-state fidelity, measuring the overlap
between ground states associated with different defect parameters,
cf. Eq.~\eqref{fiddef}. Its associated susceptibility is proportional
to the quantum Fisher information, which quantifies the reachable accuracy
of the varying defect parameter.
The fidelity provides information on the structure
change of the ground state under variations of the defect, whether
they give rise to substantial changes involving the whole system, or
the changes remain limited to a finite region. 
In particular,
within the critical crossover regime, the fidelity
susceptibility diverges as a power $L^{\varepsilon}$, where
$\varepsilon = 2y_\kappa=7/4$ for bulk defects ($y_\kappa=7/8$ is the
RG dimension of the defect parameter) and $\varepsilon = 2y_\zeta=1$
for boundary defects ($y_\zeta=1/2$ is the RG dimension of the
parameter associated with the boundary defects).  On the other hand,
the fidelity susceptibility remains finite, i.e.  $O(1)$, outside the
critical crossover region (where $|\kappa|\sim L^{-y_\kappa}$ or
$|\zeta|\sim L^{-y_\zeta}$), i.e. for defects with finite
strength. This means that a change of the defect strength, from
$\kappa>0$ to $\kappa+\delta\kappa$ with $\delta\kappa\ll 1$, causes
only local changes of the ground state, unlike at $\kappa=0$ where the
changes involve the whole critical system, giving rise to the
divergence of the fidelity susceptibility.

We have also presented numerical computations, obtained by exact
diagonalization and DMRG, to support the theoretical FSS framework
describing the critical crossover phenomena driven by
symmetry-breaking defects at CQTs. They nicely confirm the scaling
theory of the critical defect crossover.

It is worth noting that the critical crossover phenomena driven by
symmetry-breaking defects are analogous to the critical crossovers
between different fixed points of the bulk theory, arising from a
relevant perturbation at an unstable fixed point, driving away the RG
flow toward a stable fixed point (see, e.g., Refs.~\cite{PRV-98,PV-02}).

The critical crossover phenomena driven by defects that we have
discussed in this paper are quite general. Their emergence is expected
to occur in generic models at CQTs in the presence of
symmetry-breaking defects. Analogous phenomena are expected in the
presence of $n>1$ defects located in the bulk.  Of course, for
higher-dimensional systems another relevant factor concerns the
spatial dimension of the defect, i.e., if it is localized at one point,
along a line or within a surface. However, the RG arguments outlined
here can be straightforwardly extended to allow for more complex
structures of defects in higher-dimensional quantum models. In this
respect, one key feature is related to the value of the RG dimension
of the corresponding defect parameter.

For example, one may consider the effect of one symmetry-breaking
defect, such as that in Eq.~(\ref{dxdef}), in the bulk of a
two-dimensional quantum Ising model with symmetric boundary
conditions, such as PBC or OBC.  Then, one can analyze the
corresponding RG perturbation (\ref{pft}) using the RG dimensions
associated with the three-dimensional Ising universality class (see
e.g. Refs.~\cite{PV-02,RV-21} and references therein), obtaining a
positive RG dimension $y_\kappa$ for the defect parameter,
$y_\kappa = z - y_\varphi = (z-\eta)/2 \approx 0.482$ (using $z=1$ and
$\eta\approx 0.036$). One then expect that one single
symmetry-breaking defect in (strictly symmetric) two-dimensional
quantum Ising system drives a critical crossover analogous to that
found in quantum Ising rings, characterized by the divergence of the
fidelity susceptibility associated with the defect (in this case we
again expect $A_2\sim L^{2 y_\kappa}$ within the critical defect
crossover, with $2 y_\kappa\approx 0.964$).

Finally, the quantum-to-classical
mapping~\cite{Sachdev-book,RV-21} allow us to extend the quantum
scaling scenarios to classical systems with one more spatial dimension
and also one more dimension of the defect.  For example, a critical
crossover scenario analogous to that driven by one symmetry-breaking
defect in critical quantum Ising rings is expected to emerge in the
case of a classical two-dimensional Ising systems defined in a slab
with PBC and in the presence of a defect line that breaks the
${\mathbb Z}_2$ symmetry.

We conclude by mentioning that the scaling theory put forward in this paper,
which describes the critical crossover behaviors driven by symmetry-breaking
defects, can be verified with high accuracy in spin systems
with a few dozen of qubits
[as numerically done for chains of length $L \sim O(10)$].
It would be tempting to check our predictions in near-term experiments
with quantum simulators operating on a limited amount of controllable
qubits~\cite{Simon-etal-11, Debnath-etal-16, Cervera-18}.

\appendix

\section{DMRG computations on chains with PBC}
\label{dmrg}

DMRG is commonly used with OBC rather than PBC,
as the former allows us to obtain more accurate numerical results, a fact that is
usually associated with the presence of entanglement entropy in the
system at hand. Nevertheless, we used DMRG with PBC by enforcing the
presence of the additional operator
\begin{equation}
  -\hat \sigma^{(1)}_1 \otimes \hat{\mathbb{1}} \otimes \ldots
  \otimes \hat{\mathbb{1}} \otimes \hat \sigma^{(1)}_L
\end{equation}
into the superblock Hamiltonian.

We have implemented a combination of the standard two-site infinite-system
and finite-system DMRG algorithms (resp.~iDMRG and fDMRG)~\cite{W-92, Sc-11}.
Before computing any observable in the fDMRG, the stability of the
ground-state energy is verified within a discrepancy of $10^{-8}$ between
two sequential lattice sweeps.
We have also checked the stability of all our outcomes under increasing
the bond dimension $m$ of the blocks. Data for $\kappa=1, 10$, at the quantum
critical point, show small errors
(smaller than the marker size, in all the plots presented in this paper)
for a relatively small bond dimension $m=15, 12$, for
$L=32$ and $40$, respectively.

\end{document}